# P-orbital spin generator with large spin Hall angle and long spin diffusion length

*Gen Li[1†], Ying Zhang[2,3†], Xiaoguang Xu[1*], Lei Shen[4,3*], Zheng Feng[5], Kangkang Meng[1], Ang Li[6], Lu Cheng[1], Kang He[5], Wei Tan[5], Yong Wu[1], Yihong Wu[2,3], Yong Jiang[1,7*]*

[1]*Key Laboratory of Advanced Materials and Devices for Post-Moore Chips, Ministry of Education, School of Materials Science and Engineering, University of Science and Technology Beijing; Beijing 100083, China.*

[2]*Department of Electrical and Computer Engineering, National University of Singapore; 117576 Singapore.*

[3]*National University of Singapore (Chong Qing) Research Institute; Chongqing Liang Jiang New Area, Chongqing 401123, China.*

[4]*Department of Mechanical Engineering, National University of Singapore; 117575 Singapore.*

[5]*Microsystem & Terahertz Research Center, CAEP; Chengdu 610200, China.*

[6]*Faculty of Materials and Manufacturing, Beijing Key Lab of Microstructure and Properties of Advanced Materials, Beijing University of Technology; Beijing 100124, China.*

[7]*Institute of Quantum Materials and Devices, School of Electronic and Information Engineering; State Key Laboratory of Separation Membranes and Membrane Processes, Tiangong University; Tianjin 300387, China.*

*Corresponding author. Email: xgxu@ustb.edu.cn (X.G.X.); shenlei@nus.edu.sg (L.S.); yjiang@ustb.edu.cn (Y.J.)

†These authors contributed equally to this work.

**Abstract**

High density data storage and spin-logic devices require highly efficient all-electric control of spin moments. So far, charge-to-spin conversion through the spin Hall effect (SHE) highly limits to d-orbital materials associated with strong spin-orbit coupling (SOC), especially heavy metals. However, d-orbital heavy metals with strong SOC results in a short spin diffusion length, which restricts the spin transport and accumulation in spintronic devices. Therefore, it is urgent to discovery new SHE materials with both large spin Hall conductivity and high spin transport ability beyond d-orbital materials. Here, we experimentally report a large charge to spin conversion in a p-orbital $In_2Bi$ alloy, exhibiting the coexistence of a large spin Hall angle and a long spin diffusion length (4 times that of Pt). Our first-principles calculations reveal that small gap openings near the Fermi level lead to large Berry curvature-related spin Hall conductivity. Due to the delocalized nature of p-orbitals of $In_2Bi$, its spin current can overcome the physical barriers between spin Hall angle and spin diffusion length in d-orbital metals, thereby advancing the development of high performance spintronic devices.

## INTRODUCTION

Highly efficient charge-to-spin conversion is a major scientific issue for future spintronic devices. The spin Hall effect (SHE) is a well-known source of spin current by separating spin electrons to opposite material surfaces. Conventionally, d-orbital heavy metals (HM) enable the SHE due to their strong spin-orbit coupling (SOC)(*1-3*).

Upon injection into adjacent ferromagnetic (FM) layers, these spin currents induce spin-orbit torque (SOT) to switch the magnetization of the FM layer (*4-9*). Therefore, the spin current generation and its transportation within HM/FM interface dominant the performance of the SOT-based spintronic devices (*10-14*).

Traditionally, heavy 5d transition metals with strong SOC and large spin Hall angle (SHA, $\theta_{SH}$), such as Ta (*15, 16*), W (*17-20*) and Pt (*21-24*), are primary spin current sources. However, the localized d-orbital and strong SOC scattering to electrons result in short spin diffusion length (SDL, $\lambda$) in these heavy metals. This significantly restricts the effective propagation of spin currents, as illustrated in Figure 1a, and their spintronic applications.

The intrinsic Berry curvature plays a key role in spin current generation via SHE(*25-28*), which induces non-uniform electron motion in the energy bands, resulting in transverse spin accumulation. A spin current source material based on SHE usually has a large Berry curvature and strong SOC. For example, SHE of Pt is significant (*29, 30*) while those of Al and Cu are extremely weak (*31*). The different SHE behaviors have been attributed to SOC strength among these metals. However, the spin Hall conductivity (SHC) deduced from Berry curvature is not only related to the SOC strength(*32-34*), but also depends on the band structure of materials(*35*). Therefore, considering the restriction of strong SOC on $\lambda$ in d-orbital materials, it is desirable for SOT-driven spintronic devices to explore spin current sources with large Berry curvature but moderate SOC only containing p-orbital elements.

In this study, a p-orbital $In_2Bi$ alloy is proved to be a spin current source material via SHE both theoretically and experimentally (Figure 1b). Our spin transport experiments show that the SHA of $In_2Bi$ is comparable to that of Pt, while its SDL is 4 times as much as that of Pt. First-principles calculations reveal that $In_2Bi$ has delocalized p-bands near the Fermi level and a large Berry curvature, which accounts for the coexistence of large SHA and long SDL. The observation of p-orbital spin current in $In_2Bi$ overcomes the physical limitation of spin current sources in heavy 5d transition metals and expands the possibilities for designing new SOT-based spintronic devices.

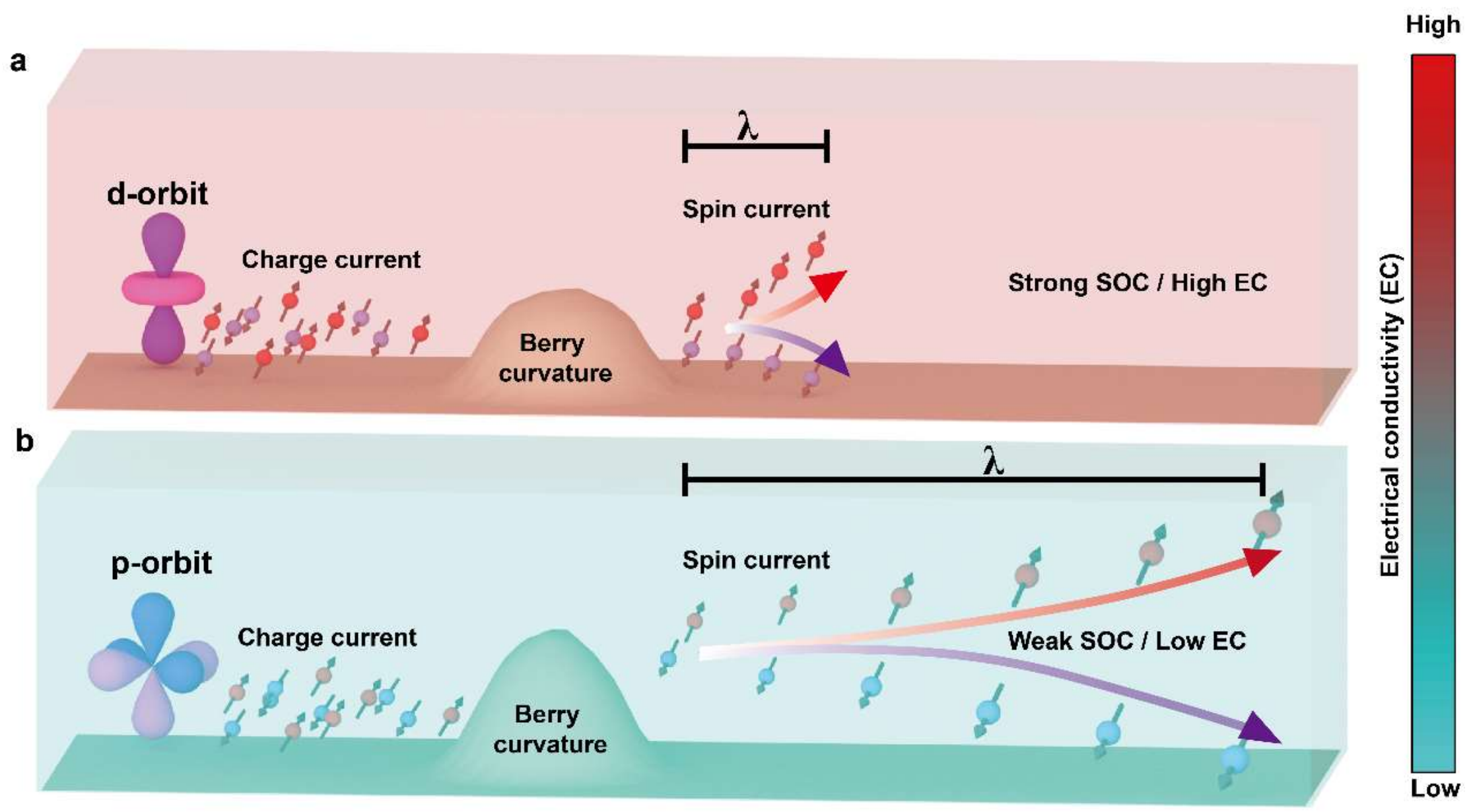

**Figure 1. Physical mechanisms of d-orbital and p-orbital materials. a**, Traditional d-orbital spin current source materials and **b**, p-orbital spin current source materials. The color bar represents the electrical conductivity (EC) levels of materials.

## RESULTS

### $In_2Bi$ synthesis and characterization

$In_2Bi$ films with thicknesses ranging from 3 to 40 nm were deposited on $Si/SiO_2$ substrates by using magnetron sputtering. The roughness of the $In_2Bi$ film is tested by atomic force microscopy (AFM). The 2D and 3D images of a 13 nm $In_2Bi$ and 3 nm Co films shown in Figure S1 reveals that $In_2Bi$ grows into a granular film with a root mean square roughness $R_q$ of approximately 2.50 nm, and Co with a roughness $R_q$ of 0.28 nm. Moreover, the $In_2Bi$ film is test to be insulated, due to the granular character. It is worth noting that although the $In_2Bi$ film exhibits insulating behavior due to its granular structure, its electrical conductivity can be verified using a parallel-resistor model, which enables the extraction of the film resistivity as a function of thickness. Further details are provided in Note S1. Grazing incidence X-ray diffraction (GI-XRD) pattern presented in Figure 2a is well consistent with the standard PDF card of $In_2Bi$ (PDF: 71-227), revealing good crystallinity of the film. Meanwhile, the enhanced diffraction intensity of (102) peak indicates a preferred orientation of the film. Considering the potential sensitivity of measurements to interface roughness, $In_2Bi$ is deposited on the Co layer to achieve a uniform and smooth $In_2Bi$/Co interface for the spin transport measurements.

X-ray photoelectron spectroscopy (XPS) measurements are conducted to decide the binding state of the $In_2Bi$ and the XPS spectra for In and Bi elements were calibrated against carbon. Figure 2b shows the XPS spectrum of the outer-shell orbitals of In and Bi near the Fermi level, together with the fitting curves. The peaks observed within the binding energy range of -5 eV to 35 eV correspond to 5p and 4d orbitals of In, as well as the 6p, 6s, and 5d orbitals of Bi. Figure 2c provides a magnified view of the binding energy region from -2 eV to 12.5 eV, highlighting that the 5p orbital of In and the 6p orbital of Bi locate close to the Fermi level. Comparing to d-orbital, p-orbital of In and Bi take on an obviously delocalized character. The In 3d and Bi 4f core level spectra (Figure 2d and 2e) indicate that the film is composed by In and Bi elements only.

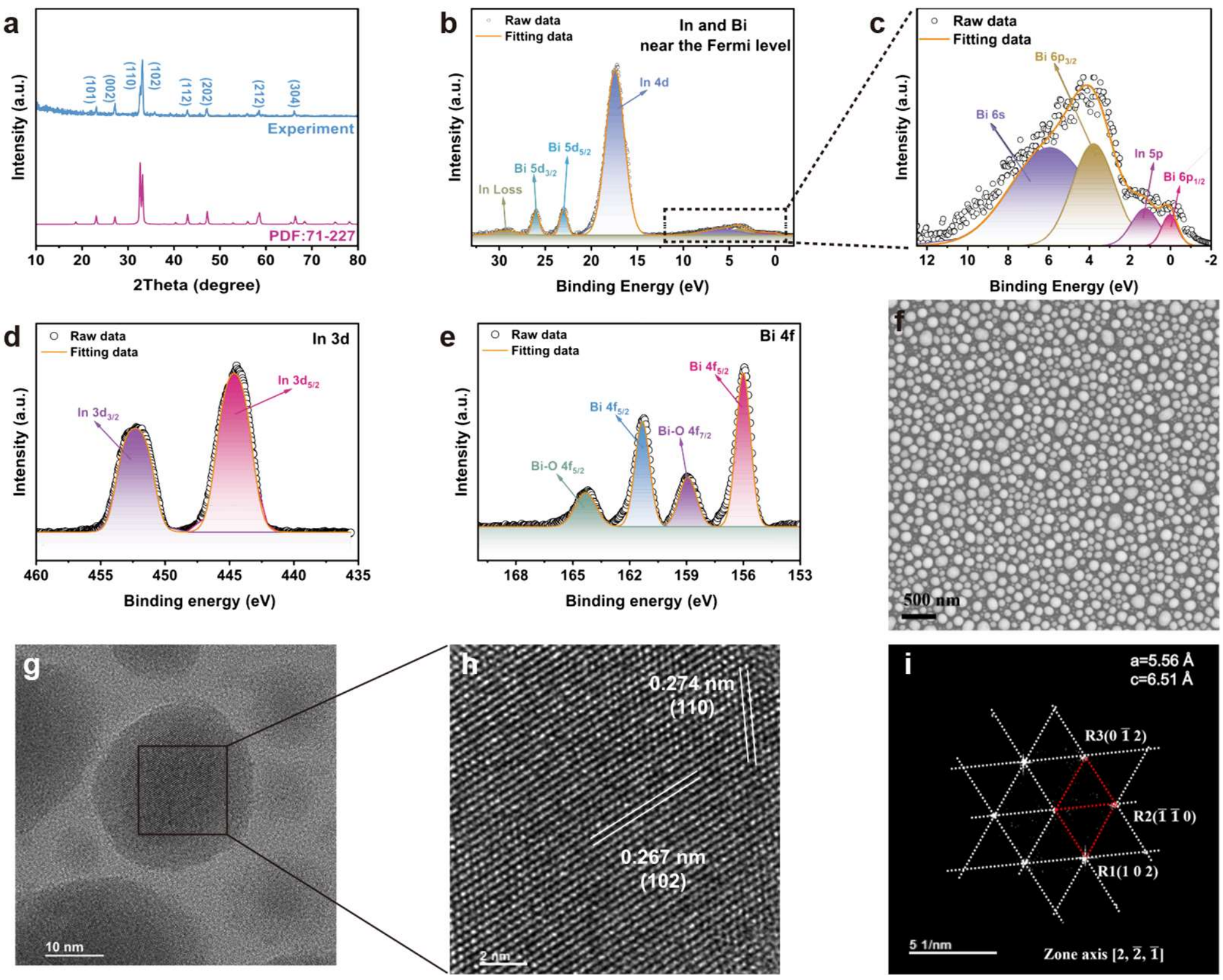

**Figure 2. Composition and crystallization structure of $In_2Bi$ thin films. a**, XRD pattern of $In_2Bi$ film grown on $Si/SiO_2$ substrate, together with PDF card (number: 71-227). **b**, XPS spectrum for Bi and In near the Fermi level in $In_2Bi$. **c**, XPS spectrum of the binding energy range from -5 eV to 12.5 eV. **d**, In 3d and **e**, Bi 4f XPS spectra of $In_2Bi$. **f**, SEM image of 13 nm $In_2Bi$ film. **g**, TEM

image of 13 nm $In_2Bi$ single particle, **h**, enlarged TEM image and **i**, selected area diffraction pattern of the corresponding area.

The scanning electron microscope (SEM) image in Figure 2f clearly shows that the $In_2Bi$ film grown on the $Si/SiO_2$ substrate exhibits a granular morphology, consistent with the result of AFM. Transmission electron microscopy (TEM) characterization was performed on $In_2Bi$ film and the images are presented in Figure 2g and Figure 2h. The granular film is composed by single crystals, with the size ranging mainly from 30 to 40 nm. Figure 2h shows good lattice structure of a single particle which can be indexed to (102) and (110). The diffraction pattern in Figure 2i can be indexed into (102), $(\bar{1}\bar{1}0)$, and $(0\bar{1}2)$, indicating a hexagonal crystal structure of $In_2Bi$. To exclude the possibility that the observed thickness dependence arises from structural or crystalline variations of $In_2Bi$, we performed TEM imaging and corresponding Fast Fourier Transform analyses on 25 nm and 40 nm granular $In_2Bi$ films (Figure S2 a–d). These results reveal identical diffraction features and lattice spacings with the 13 nm film (Figures. 2g and 2i), confirming that films of different thicknesses share the same crystalline phase. Furthermore, XRD measurements for films of varying thickness (25 and 40 nm; Figure S2e) exhibit the same diffraction peaks without additional reflections, providing further evidence for the absence of thickness-induced phase transitions. While minor variations in grain size or surface roughness are expected in granular films, no structural changes that could qualitatively alter spin transport have been observed. According to the TEM and XRD results, the $In_2Bi$ particles prefer to grow along (102) plane during sputtering.

Energy dispersive spectroscopy (EDS) mapping analysis in Figure S3 further demonstrates the content of In and Bi in the film. Moreover, the stoichiometry of the film is defined to be $In_2Bi$ by semi-quantitative element content comparison based on the $In_2Bi$ XPS spectra and quantitatively inductively coupled plasma optical emission spectrometer (ICP-OES) (See Note S2 and Figure S4). When the thickness increases, the $In_2Bi$ film transforms from granular film into continuous film. The temperature-dependent resistivity (ρ-T) (Figure S5) and magnetoresistance (Figure S6) of the 80 nm $In_2Bi$ film were measured to characterize the fundamental electrical transport properties.

**Spin Hall angle and spin diffusion length measurements**

Spin Hall magnetoresistance (SMR) measurements are employed to examine the spin current generation and diffusion in $In_2Bi$, as shown in Figure 3a. The influence of the Co/$In_2Bi$ interface alloying on the SMR is excluded by sweeping field tests as shown in Figure S7. The specific mechanism of SMR testing is shown in Note S3. Specifically, Hall bar devices were fabricated on a Si/$SiO_2$ substrate with the stacking structure of Co (3 nm)/$In_2Bi$ (t)/$Al_2O_3$ (3 nm) (thickness in nanometer, t=0, 3, 5, 7, 9, 11, 13, 15, 17, 25, and 40). The SMR curve for each $In_2Bi$ thickness is shown in Note S4 and Figure S8, together with fitting curves. As summarized in Figure 3b and Figure 3c, the SMR signal increases gradually with the thickness of $In_2Bi$ and reaches the maximum of approximately 0.093% when t = 13 nm. Subsequently, the signal decreases with the thickness further increasing. The relationship between the thickness and the extracted SMR signal for the FM/NM heterojunction can be expressed as follows:

$$\frac{\Delta R_{XX}^{SMR}}{R_{XX}^{0}} \sim -\theta_{SH}^{2}\frac{\lambda_N}{t_N}\frac{tanh^2\left(\frac{t_N}{2\lambda_N}\right)}{1+\xi}\times\left[\frac{g_R}{1+g_R\coth\left(\frac{t_N}{\lambda_N}\right)}-\frac{g_F}{1+g_F\coth\left(\frac{t_N}{\lambda_N}\right)}\right]$$

$$g_R \equiv 2\rho_N\lambda_N Re[G_{MIX}]\ ,\ g_F \equiv \frac{(1-P^2)\rho_N\lambda_N}{\rho_F\lambda_F\coth\left(\frac{t_F}{\lambda_F}\right)} \tag{1}$$

Here, $R^0$ denotes the resistance when the magnetization is aligned along the Z-axis. The variables $t_N$、$\rho_N$、$\lambda_N$, and $\theta_{SH}$ represent the thickness, resistivity ($6.17\times10^{-5}$ $\Omega\cdot cm$, extracted from Figure S5.), spin diffusion length, and spin Hall angle of the non-magnetic layer, respectively. $G_{MIX}$ refers to the spin mixing conductance. Additionally, $t_F$, $\rho_F$, $\lambda_F$, and P correspond to the thickness, resistivity ($7.5\times10^{-6}$ $\Omega\cdot$cm) (*36, 37*), SDL (10 nm) (*38*), and spin polarization (0.34) (*39*) of the ferromagnetic layer, respectively. The parameter $\xi$ describes the shunting effect entering the magnetic layer and can be expressed as $\frac{\rho_{NM}d_{FM}}{\rho_{FM}d_{NM}}$.

We employed Equation (1) to analyze the variation trend of SMR in the Co (3 nm)/ $In_2Bi$ (t)/$Al_2O_3$ (3 nm) heterojunction. The fitting results are well consistent with experimental data for different $In_2Bi$ thickness. Meanwhile, the $\theta_{SH}$ and $\lambda$ of $In_2Bi$ are defined to be 0.106 and 3.68 nm, respectively. We also performed SMR measurements for Co/Pt devices and provided the results in Figure S9. The SMR fitting results yield a spin Hall angle of 0.136 and a spin diffusion length of 1.175 nm, which are consistent with previously reported SMR studies on Pt (*40*). According to the SMR results, the $\theta_{SH}$ of $In_2Bi$ alloy is comparable to that of Pt, but the $\lambda$ is about 4 times as much as that of Pt (*41, 42*), Therefore, $In_2Bi$ is a potential spin generator with higher spin current transport ability (*43*).

The terahertz (THz) emission spectroscopy is launched to systematically investigate the spin-to-charge conversion in $In_2Bi$ besides electrical transport measurements. The mechanism of the THz emission is illustrated in Figure 3d and Note S5. The thickness of Co layer is fixed at 3 nm, while the $In_2Bi$ layer thickness varies from 3 nm to 40 nm, as shown in Figure 3e. We quantified the THz signal amplitude by calculating the difference between the peak and valley values of their corresponding waveforms, as summarized in Table S1. The results demonstrate that the THz signal amplitudes of the Co/$In_2Bi$ heterojunction initially rises and then declines with the thickness of $In_2Bi$ increasing, as shown in Figure S10a. Notably, similar to the observations in SMR, the THz signal amplitude also peaks at the $In_2Bi$ thickness of 13 nm (as shown in Figure S10a).

To assess the THz measurements in Co/$In_2Bi$ bilayers, Co/Pt heterostructures are employed as comparation, which results are shown in Note S6, Figure S10b and Figure S11. The polarization of the THz waveforms for both Co/$In_2Bi$ and Co/Pt bilayers is the same, indicating that $In_2Bi$ and Pt exhibit the same direction of spin current. Moreover, the THz polarization of Co (3 nm)/W (3 nm) stack shown in Figure S12 further supports that the spin current direction of $In_2Bi$ is consist with Pt but opposite to W.

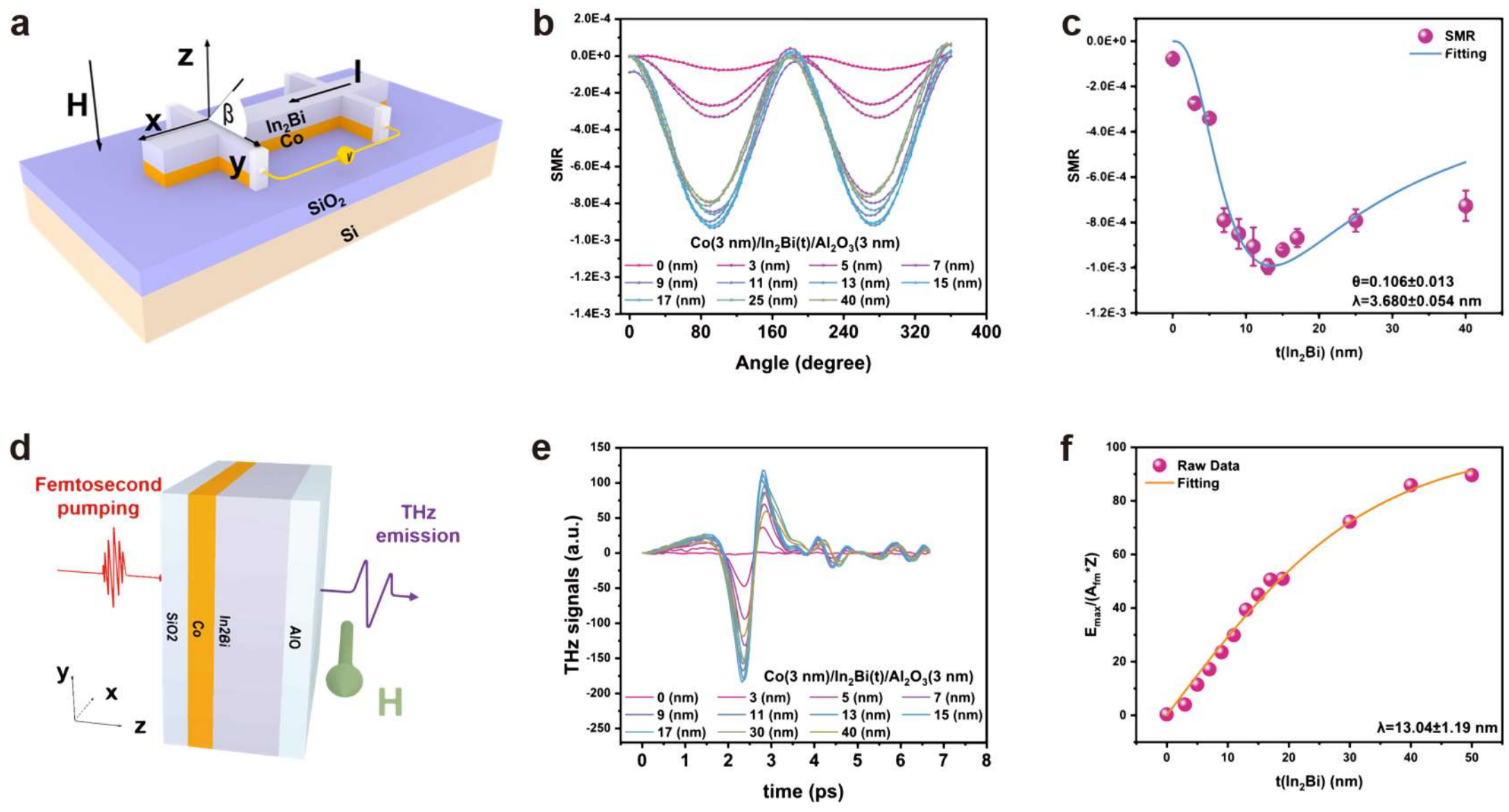

**Figure 3. The spin Hall magnetoresistance and THz emission measurement results of Co/$In_2Bi$ heterostructures together with fitting curves.** **a**, Schematic diagram of heterostructure for spin Hall magnetoresistance measurement. **b**, Angle-dependent spin magnetoresistance curves of Co (3 nm)/$In_2Bi$ (t)/$Al_2O_3$ (3 nm) with varying $In_2Bi$ thickness measured under 9 T at room temperature. **c**, Dependence of the SMR amplitude on the $In_2Bi$ thickness in Co (3 nm)/$In_2Bi$ (t)/$Al_2O_3$ (3 nm) heterojunction and fitting curve. **d**, Schematic illustration of THz emission measurement geometry. The femtosecond laser propagates along z axis and a magnetic field is applied along x axis. **e**, THz signals for Co (3 nm)/$In_2Bi$ (t)/$Al_2O_3$ (3 nm) multilayers. **f**, The $In_2Bi$ thickness dependence of the normalized THz amplitude for Co (3 nm)/$In_2Bi$ (t)/$Al_2O_3$ (3 nm) THz emitter. The symbol and line reveal the experimental and fitting results, respectively.

Figure 3f depicts the relationship between THz signal amplitude and layer thickness for Co/$In_2Bi$/$Al_2O_3$ multilayers. To enable comparison, the impact of thickness variations on the laser absorption of FM layer ($A_{FM}$) and the THz emission impedance (Z) is evaluated. The theoretical fitting curve shown in Figure 3f are derived from the following equation (*44, 45*):

$$E_{THz} = A_{FM} \cdot Z \cdot G \cdot \theta_{SHE} \cdot \left[tanh\left(\frac{t_N}{2\lambda}\right) \cdot \lambda\right]. \quad (2)$$

Here, $E_{THz}$ denotes the amplitude of the measured THz signals, $A_{FM}$ is the laser absorption of the FM layer, Z is the THz radiation impedance, G is a parameter related to the generation and transmission of ultrafast spin current, and $\theta_{SHE}$, $\lambda$, and $t_N$ represent the spin Hall angle, the SDL and the thickness of the NM layer, respectively. $A_{FM} = A_{total} \cdot \frac{t_F}{t_F + t_N}$, where Atotal are the total laser absorbance of the bilayer, and $t_{FM}$ is the thickness of the FM layer. By measuring the total laser absorbance, one can obtain $A_{FM}$. $Z = \frac{Z_0}{1+n+Z_0 \cdot (\sigma_F t_F + \sigma_N t_N)}$, which can be obtained by measuring the THz transmission of the sample and the bare substrate.

In equation (2), λ and $G \cdot \theta_{SHE}$ are adjustable parameters in fitting, while A and Z can be determined experimentally. The fitting curve presented in Figure 3f reveals a strong consistency between the fitted and measured results (*46*). The fitting result indicates that $In_2Bi$ has a long SDL of $\lambda_{In_2Bi}$=13.04±1.19 nm. The SDL in Pt has also been studied and simulated to be 2.93±0.24 nm, which is consistent with previous reports (*47, 48*). Furthermore, the $\lambda_{In_2Bi}$ is more than 4 times of $\lambda_{Pt}$ according to THz measurements, which is in close agreement with the results from SMR tests. It should be noted that the SHA and SDL of the same material can vary significantly depending on the experimental method. For instance, as summarized in Table S2, the reported SHA and SDL values for Pt differ widely across different measurement techniques. Nevertheless, the SHA and SDL values we obtained for Pt using SMR and THz emission methods are similar with those previously reported by corresponding techniques, confirming the reliability of our analysis. Although the absolute values of λ are different for SMR and THz fitting results due to the different mechanism (*42, 45*), the ratio of $\lambda_{\mathrm{In_2Bi}}/\lambda_{Pt}$ obtained from the same method can reflect the capability of spin transport of materials. Therefore, $In_2Bi$ possesses a significant long SDL comparing to transition metals with strong SOC, which is beneficial for $In_2Bi$ as a spin generator.

**Density functional theory calculations**

To further investigate the origin of large SHA and long SDL in hexagonal $In_2Bi$, we perform first-principles calculations. The $In_2Bi$ crystal has a hexagonal $P6_3/mmc$ space group (No. 194) with lattice parameters a = b = 5.559 Å and c = 6.508 Å (Figure 4a), and its Brillouin zone is shown in Figure 4b. The schematic and calculated band structures in the absence and presence of SOC are shown in Figure 4c and 4d, respectively. As can be seen, in the absence of SOC, there are several band crossings near the Fermi level. These band openings are from the influence of a moderate effective SOC originating from the combined contributions of Bi and In p orbitals (Figure 4e) with In p-states dominating near the Fermi level. Overall, this leads to a moderate SOC that is weaker than that of pure Bi metal, yet sufficient to induce observable band splitting. It is worth noting that the spin Hall conductivity first increases and then decreases with increasing SOC (*49, 50*). This explains why Bi metal has a larger SOC than Pt but smaller SHC. Therefore, in the search for high performance p-orbital SHE materials, the key physics lies in a moderate SOC as well as how SOC modifies the band structure and generates Berry curvature near the Fermi level. In $In_2Bi$ (Figure 4f), there are several small openings within the energy range of ±0.5 eV, stabilizing the nonzero components of SHC (the color changes with opposite signs in Figure 4f). Figure 4g shows three nonzero tensor elements of SHC plotted as a function of energy. The integral SBC of occupied states along high-symmetry paths (Figure 4h) reveals a maximum and a minimum near the Γ-point and H-point, respectively, perfectly aligning with the projected spin Berry curvature on the band structure (Figure 4f).

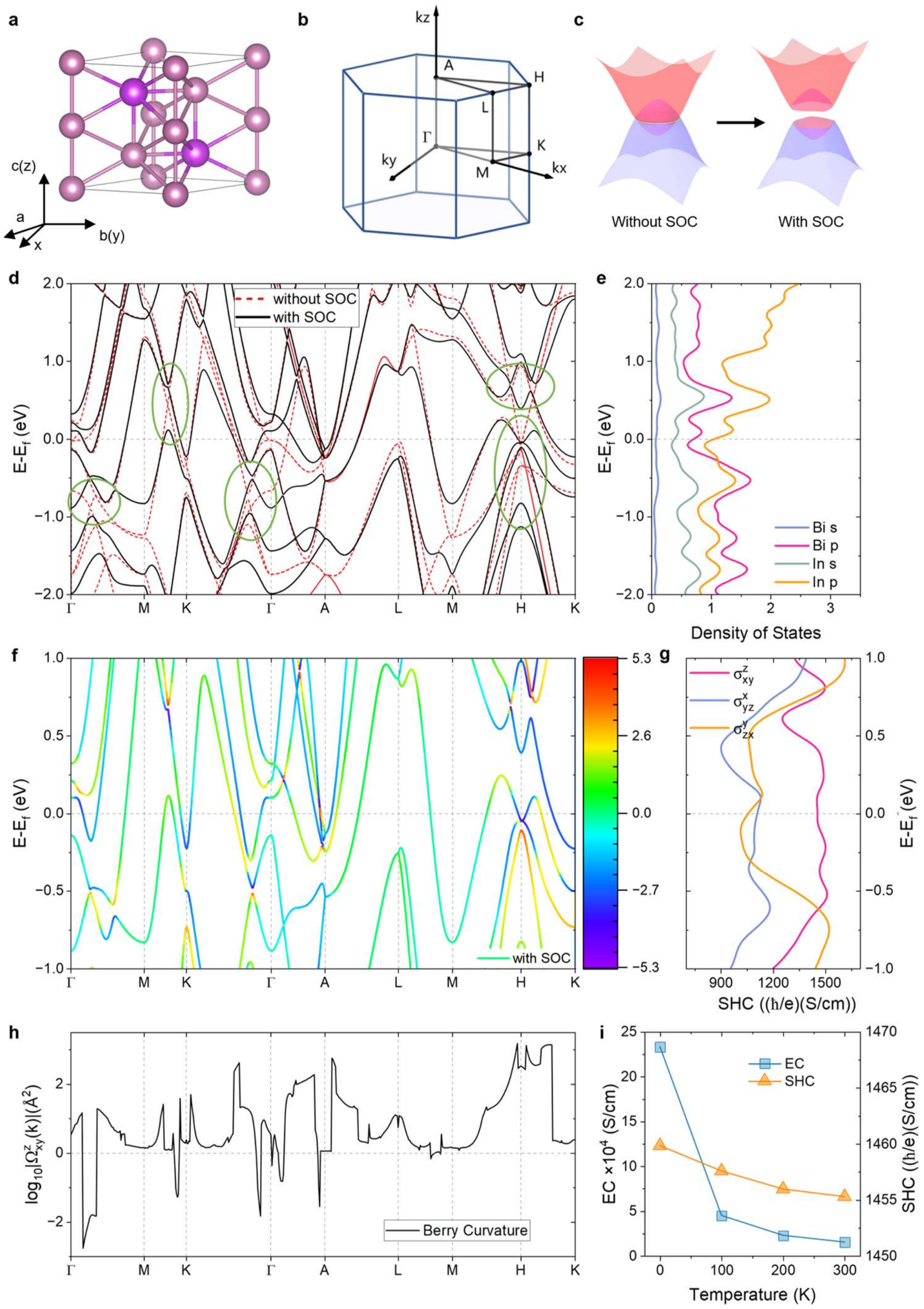

**Figure 4. Crystal and electronic structures of $In_2Bi$, and spin Hall conductivity, electrical conductivity and spin Berry curvature. a**, $In_2Bi$ crystal structure with space group $P6_3/mmc$ (194). The a, b, c are the crystal-axis directions. **b**, Brillouin zone for $In_2Bi$. **c**, Schematic of the band splitting under SOC, and further contributes to large SHC. **d**, The calculated band structures along the high symmetry directions with (black) and without (red) SOC. The green circles highlight the opening gaps by SOC near the Fermi level. **e**, The projected density of states. **f**, The band structure of $In_2Bi$ projected by spin Berry curvature of $\Omega_{xy}^{z}$, where the red (blue) color denotes a positive (negative) contribution of the spin Berry curvature in units of Å$^2$. **g**, Three nonzero tensor elements of SHC plotted as a function of energy. **h**, The k-resolved spin Berry curvatures of $\Omega_{xy}^{z}$ on a log scale. **i**, The temperature dependence of spin Hall conductivity (orange) and electrical conductivity (blue) at the Fermi energy.

The SHA is defined by the ratio of the SHC, $\sigma_{xy}^{z}$ the spin current $J_s$ in *x* direction with spin polarization along *z* direction due to an incoming charge current from the *y* direction, to the longitudinal electrical conductivity (EC), $\sigma_{xx}$, i.e. $\theta_{SH} = \frac{2e}{\hbar}|\frac{\sigma_{xy}^{z}}{\sigma_{xx}}|$, which shows that EC also plays an important role on the magnitude of SHA. Thus, we further investigate the temperature-dependent variations of both EC and SHC, as shown in Figure 4i. The calculated SHC of $In_2Bi$ is approximately 65% that of Pt(*51*). However, $In_2Bi$ exhibits relatively low electrical conductivity (with a theoretical EC value only one-fifth that of Pt (*52*). This combination results in a theoretical SHA for $In_2Bi$ of 0.15 at room temperature, which is about two times as much as that of Pt(*53*).

It is worth noting that our experimental SHA is smaller than that of theoretical value. This can be explained by the crystal orientation of the film. The XRD and TEM analysis reveal that there are two dominated (102) and (110) planes in the $In_2Bi$ layer. We thus further calculate the SHC of these two planes (Note S7, Figure S13 and Figure S14), where the $\sigma_{xy}^{z}$ of (102) and (110) planes are around 517 and 674 $\hbar/e(S/cm)$, respectively, which show a decrease compared with the bulk phase (about 1455 $\hbar/e(S/cm)$) and account for the difference between experimental and theoretical SHA values.

## DISCUSSION

Based on the experimental and theoretical results, due to the large Berry curvature and the relatively low basic conductivity, $In_2Bi$ possesses a large spin Hall angle and a long spin diffusion length, which makes it capable of efficient charge-to-spin conversion. As illustrated in Figure 1a, traditional d-orbital spin current source materials, such as heavy metals, depend on the strong SOC originating from 5d elements. Such strong SOC contributes to the large SHA. However, the localized DOS of d-orbital and the large scattering by strong SOC lead to a small SDL, confining the spin current propagation. To overcome such a dilemma in d-orbital metals, we demonstrate that $In_2Bi$, an IIIA-VA group p-orbital alloy, exhibits both large SHC and long SDL (Figure 1b). Owing to the large Berry curvature and the delocalized character of the p-orbital, $In_2Bi$ alloy enables the spin current propagation, as well as effective charge-to-spin conversion. Therefore, p-orbital materials like the $In_2Bi$ alloy can also serve in charge-to-spin conversion, breaking the limitation of conventional 5d heavy materials.

Notably, although the $In_2Bi$ films in this work takes on a granular morphology, they still generate and transport spin current efficiently. The SDL of $In_2Bi$ should be decided by both intrinsic property and Elliott–Yafet mechanism. The spin scattering due to morphology and crystallin size will greatly restrict the experimental SDL. Very recently, a van der Waals squeezing method has been reported to realize single crystal films of low-melting-point metals(*54*). Although precise thickness control is challenging with this technique, Co/$In_2Bi$ heterostructures with continuous $In_2Bi$ layers are successfully obtained. AFM characterization, confirms a substantial improvement in surface morphology after hot pressing. Correspondingly, these hot-pressed devices exhibit robust SMR signals, substantially larger than that of the Co (3 nm)/$Al_2O_3$ (3 nm) control device. More importantly, the SMR signal is significantly enhanced after hot pressing. These findings collectively confirm that the observed SMR originates from spin-to-charge interconversion in $In_2Bi$ itself, rather than spurious contributions from granularity. More details are shown in Note S1 and Figure S16. One can expect $In_2Bi$ to have high performance closing to theoretical properties in devices if high-quality single crystalline films were realized by advanced techniques.

In summary, we have experimentally and theoretically demonstrated that $In_2Bi$ is an efficient p-orbital spin source material that overcomes the conflict between SHA and SDL commonly observed in conventional d-orbital heavy metals. SMR and THz emission spectroscopy consistently reveal that $In_2Bi$ exhibits a SHA comparable to that of Pt, while maintains a significantly longer SDL. First-principles calculations attribute this coexistence of large SHA and long SDL to the moderate SOC and large Berry curvature arising from delocalized In and Bi p-orbitals. Most importantly, $In_2Bi$ establishes a new mechanism for spin source design, demonstrating that p-orbital alloys can achieve efficient charge-to-spin conversion without relying on the strong SOC of d-orbital metals. These findings not only identify $In_2Bi$ as a promising spin current generator but also broaden the material design strategies for next-generation spintronic devices.

## MATERIALS AND METHODS

### Magnetron sputtering

Magnetron sputtering leverages the interplay between magnetic and electric fields to exert precise control over plasma dynamics, effectively enhancing both the sputtering rate and film quality. Based on the principles of physical vapor deposition (PVD), magnetron sputtering relies on high-energy ions to bombard the target material, resulting in the ejection of atoms from the target surface, which then deposit onto a substrate to form a thin film.

The base pressure for thin film deposition was maintained at $9\times10^{-6}$ Pa. For the deposition of the $In_2Bi$ layer, sputtering was performed at an Ar pressure of 0.6 Pa with a DC power of 4 W. Co was sputtered at the same pressure (0.6 Pa) with a DC power of 40 W. For Pt, sputtering was carried out at an Ar pressure of 0.4 Pa with a DC power of 15 W. $Al_2O_3$ was deposited at an Ar pressure of 0.8 Pa using RF power set to 50 W.

### Terahertz emission

In this study, we employed a standard THz time-domain spectroscopy setup to generate and detect THz pulse waveforms. A linearly polarized femtosecond laser pulse (120 fs duration, 800 nm central wavelength, 500 mW (Pin), 80 MHz repetition rate) was incident on the thin film along the z-axis. A magnetic field of 0.1 T (B) was applied along the x-axis to orient the magnetization of the ferromagnetic layer. The THz signal, characterized by an electric field oriented along the y-axis, was emitted from the device and detected by a photoconductive antenna. All measurements were conducted in a dry air environment at room temperature. In principle, either side of the sample can be pumped—either the substrate side or the ferromagnetic layer side. However, in the current experiment, pumping was performed from the substrate side to mitigate undesired THz absorption by the substrate, thereby facilitating THz emission from the surface of the thin film.

### Density functional theory calculations

In order to calculate the spin hall conductivity, the Kubo formula in the clean limit is used for calculation,

$$\sigma_{xy}^{z} = e\hbar \int_{BZ} \frac{dk}{(2\pi)^3} \sum_n f_{nk} \Omega_{xy}^{n,k}(k) \tag{3}$$

where $f_{nk}$ is the Fermi-Dirac distribution function for the $n^{th}$ band at $k$. The SHC tensor element $\sigma_{xy}^{z}$ desctibes the spin current $J_x$ with spin polarization along z direction due to an incoming charge current from the y direction. Other elements in the third-order tensor can be obtained by changing the mutually orthogonal Cartesian directions. The $\Omega_{xy}^{n,k}(k)$ refers to the Berry curvature of the nth band is defined as:

$$\Omega_{xy}^{n,k}(k) = -\sum_{m\neq n} \frac{\left(2Im\left[\langle nk|j_x^z|mk\rangle\langle mk|v_y|nk\rangle\right]\right)}{\left((\epsilon_{nk}-\epsilon_{mk})^2 - (i\eta)^2\right)} \tag{4}$$

The first-principles calculations are carried out with QUANTUM ESPRESSO(*55, 56*) based on the projector augmented-wave (PAW) method and a plane-wave basis set. The generalized gradient approximation (GGA) and Perdew-Burke-Ernzerhof (PBE) functional are selected to deal with the electron exchange-correlation interactions. Before calculation, the structure was fully relaxed with the force on each atom was less than $10^{-3}$ Ry/Bohr. A fully relativistic pseudopotential, implemented in the PSLIBRARY(*57*) were used to calculate the electronic band structure with SOC. The plane-wave and charge-density cutoff energy are 80 Ry and 400 Ry, respectively. A k-point grid of 10 × 10 × 10 was used in the self-consistent calculations. Spin-orbit interaction was taken into account self-consistently, which also includes scalar relativistic effects. Once the self-consistent calculations were completed, the Bloch functions were Fourier transformed to the maximally localized Wannier functions (MLWFs) using the WANNIER90 package(*58*). The chosen Wannier functions are In: s, p, Bi: s, p. After the k-point grid convergence test and method comparison (Note S8), the WannierBerri algorithm considering material crystal symmetry can converge to an accurate result under relatively coarse k-mesh with shorter calculation time. Therefore, the calculation method from Qiao et al.(*51*) integrated in the Wannierberri package(*59*) was used to complete subsequent calculations. The SHC of bulk $In_2Bi$ was calculated on a dense 100 × 100 × 100 k-mesh while for the structure of $In_2Bi$ in different phase, it was calculated on a relatively coarse 50 × 50 × 50 k-mesh using WannierBerri.

In order to evaluate the electrical conductivity (EC), the energy and k-point dependent carrier relaxation time is calculated by utilizing EPW package(*60*). Since the SOC has little effect on the force constants and phonon frequencies, we employ norm-conversing pesudopotentials(*61*) with the Perdew-Burke-Ernzerhof (PBE) form of the exchange-correlation functional for electrical conductivity calculation. The cutoff energy of the plane wave is set as 80 Ry, while the convergence threshold of electron energy is set to be $10^{-12}$ Ry for the self-consistent field calculation. The phonon dispersion relations are calculated by density functional perturbation theory(*62*), with a $6 \times 6 \times 6$ q grid and a self-consistency threshold of $10^{-14}$. Using the maximally localized Wannier functions basis(*63*), the electron-phonon matrix elements, band energies, and phonon modes are interpolated from an initial coarse grid of $12 \times 12 \times 12$ and $6 \times 6 \times 6$ electron and phonon vector girds, respectively, to dense grids of $60 \times 60 \times 60$ and $40 \times 40 \times 40$ electron and phonon vector girds, respectively.

**Acknowledgements**

**Funding:**

X.G.X. and Y.J. acknowledges the financial support from the National Key R&D Program of China (2022YFA1402602) and National Natural Science Foundation of China (Grant Nos. U24A6001, U24A6002 and U23A20548). L.S. acknowledges the financial support from Singapore MOE Tier 1 (No. A-8001194-00-00) and Singapore MOE Tier 2 (No. A-8001872-00-00). Z.F. acknowledges the financial support from National Natural Science Foundation of China (Grant No. 62027807).

**Author contributions:**

X.G.X. and L.S. conceived the project and designed the experiments; G.L. and Y.Z. wrote the manuscript; G.L. performed heterostructure growth, device fabrication, material characterization, and SMR measurements under the supervision of X.G.X., K.K.M., Y.W. and Y.J.; Y.Z. performed the calculations and theoretical analysis under the supervision of L.S. and Y.H.W.; Z.F., K.H. and W.T. performed the THz measurement and analyzed the results; A.L. helped in TEM characterization; L.C. helped in SMR measurement and device fabrication. All authors discussed and analyzed the results and commented on the manuscript.

**Competing interests:**

The authors declare that they have no conflict of interest.

**Data and materials availability:**

All data are available in the main text or the Supporting Information.

**Supplementary Materials**

**This PDF file includes:**

Notes S1-S8

Figs. S1-S17

Tables S1 and S2

Reference

## REFERENCES AND NOTES

1. M. C. Beeler, R. A. Williams, K. Jiménez-García, L. J. LeBlanc, A. R. Perry, I. B. Spielman, The spin Hall effect in a quantum gas. *Nature* **498**, 201-204 (2013).
2. I. A. Ado, A. Qaiumzadeh, R. A. Duine, A. Brataas, M. Titov, Asymmetric and Symmetric Exchange in a Generalized 2D Rashba Ferromagnet. *Phys. Rev. Lett.* **121**, 086802 (2018).
3. Z. G. Yu, Spin-Orbit Coupling, Spin Relaxation, and Spin Diffusion in Organic Solids. *Phys. Rev. Lett.* **106**, 106602 (2011).
4. R. Nakajima *et al.*, Giant spin polarization and a pair of antiparallel spins in a chiral superconductor. *Nature* **613**, 479-484 (2023).
5. Z. Zhou *et al.*, Manipulation of the altermagnetic order in CrSb via crystal symmetry. *Nature* **638**, 645-650 (2025).
6. Y. Fan *et al.*, Magnetization switching through giant spin–orbit torque in a magnetically doped topological insulator heterostructure. *Nat. Mater.* **13**, 699-704 (2014).
7. H. Wang *et al.*, Room temperature energy-efficient spin-orbit torque switching in two-dimensional van der Waals Fe3GeTe2 induced by topological insulators. *Nat. Commun.* **14**, 5173 (2023).
8. L. You *et al.*, Switching of perpendicularly polarized nanomagnets with spin orbit torque without an external magnetic field by engineering a tilted anisotropy. *Proceedings of the National Academy of Sciences* **112**, 10310-10315 (2015).
9. S. Varotto *et al.*, Room-temperature ferroelectric switching of spin-to-charge conversion in germanium telluride. *Nat. Electron.* **4**, 740-747 (2021).
10. C. H. Jin *et al.*, Imaging of pure spin-valley diffusion current in WS2/WSe2 heterostructures. *Science* **360**, 893-896 (2018).
11. P. Vaidya *et al.*, Subterahertz spin pumping from an insulating antiferromagnet. *Science* **368**, 160-165 (2020).
12. A. Manchon *et al.*, Current-induced spin-orbit torques in ferromagnetic and antiferromagnetic systems. *Rev. Mod. Phys.* **91**, 035004 (2019).
13. M. Dc *et al.*, Room-temperature high spin–orbit torque due to quantum confinement in sputtered BixSe(1–x) films. *Nat. Mater.* **17**, 800-807 (2018).
14. J. Ryu *et al.*, Efficient spin–orbit torque in magnetic trilayers using all three polarizations of a spin current. *Nat. Electron.* **5**, 217-223 (2022).
15. J. X. Li *et al.*, Spin current from sub-terahertz-generated antiferromagnetic magnons. *Nature* **578**, 70-74 (2020).
16. R. Yu *et al.*, Determination of spin Hall angle and spin diffusion length in β-phase-dominated tantalum. *Phys. Rev. Mater.* **2**, 074406 (2018).
17. J. Kim, P. Sheng, S. Takahashi, S. Mitani, M. Hayashi, Spin Hall Magnetoresistance in Metallic Bilayers. *Phys. Rev. Lett.* **116**, 097201 (2016).
18. P. Sheng, Y. Sakuraba, Y.-C. Lau, S. Takahashi, S. Mitani, M. Hayashi, The spin Nernst effect in tungsten. *Sci. Adv.* **3**, e1701503 (2017).
19. E.-S. Park *et al.*, Strong higher-order angular dependence of spin-orbit torque in W/CoFeB bilayer. *Phys. Rev. B* **107**, 064411 (2023).
20. Y. Li *et al.*, Enhancing the Spin–Orbit Torque Efficiency by the Insertion of a Sub-nanometer β-W Layer. *ACS Nano* **16**, 11852-11861 (2022).

21. J. W. Lee *et al.*, Enhanced spin-orbit torque by engineering Pt resistivity in Pt/Co/AlO structures. *Phys. Rev. B* **96**, 064405 (2017).
22. M. Fang *et al.*, Tuning the interfacial spin-orbit coupling with ferroelectricity. *Nat. Commun.* **11**, 2627 (2020).
23. M. Isasa, E. Villamor, L. E. Hueso, M. Gradhand, F. Casanova, Temperature dependence of spin diffusion length and spin Hall angle in Au and Pt. *Phys. Rev. B* **91**, 024402 (2015).
24. X. K. Xu *et al.*, Giant Extrinsic Spin Hall Effect in Platinum-Titanium Oxide Nanocomposite Films. *Adv. Sci.* **9**, 2105726 (2022).
25. M. V. Berry, Quantal phase factors accompanying adiabatic changes. *Proceedings of the Royal Society of London. A. Mathematical Physical Sciences* **392**, 45-57 (1984).
26. C. L. Kane, E. J. Mele, Quantum Spin Hall Effect in Graphene. *Phys. Rev. Lett.* **95**, 226801 (2005).
27. D. J. Thouless, M. Kohmoto, M. P. Nightingale, M. den Nijs, Quantized Hall Conductance in a Two-Dimensional Periodic Potential. *Phys. Rev. Lett.* **49**, 405-408 (1982).
28. D. Xiao, M.-C. Chang, Q. Niu, Berry phase effects on electronic properties. *Rev. Mod. Phys.* **82**, 1959-2007 (2010).
29. E. Saitoh, M. Ueda, H. Miyajima, G. Tatara, Conversion of spin current into charge current at room temperature: Inverse spin-Hall effect. *Appl. Phys. Lett.* **88**, (2006).
30. T. Kimura, Y. Otani, T. Sato, S. Takahashi, S. Maekawa, Room-Temperature Reversible Spin Hall Effect. *Phys. Rev. Lett.* **98**, 156601 (2007).
31. S. O. Valenzuela, M. Tinkham, Direct electronic measurement of the spin Hall effect. *Nature* **442**, 176-179 (2006).
32. M. I. Dyakonov, V. I. Perel, Current-induced spin orientation of electrons in semiconductors. *Phys. Lett. A* **35**, 459-460 (1971).
33. R. V. Shchelushkin, A. Brataas, Spin Hall effects in diffusive normal metals. *Phys. Rev. B* **71**, 045123 (2005).
34. H.-A. Engel, B. I. Halperin, E. I. Rashba, Theory of Spin Hall Conductivity in n-Doped GaAs. *Phys. Rev. Lett.* **95**, 166605 (2005).
35. G. Y. Guo, S. Murakami, T. W. Chen, N. Nagaosa, Intrinsic Spin Hall Effect in Platinum: First-Principles Calculations. *Phys. Rev. Lett.* **100**, 096401 (2008).
36. Y. K. Ko *et al.*, Studies of cobalt thin films deposited by sputtering and MOCVD. *Mater. Chem. Phys.* **80**, 560-564 (2003).
37. I. I. Amirov, R. V. Selyukov, V. V. Naumov, E. S. Gorlachev, Influence of Deposition Conditions and Ion-Plasma Treatment of Thin Cobalt Films on Their Electrical Resistivity. *Russ. Microelectron.* **50**, 1-7 (2021).
38. G. Zahnd *et al.*, Spin diffusion length and polarization of ferromagnetic metals measured by the spin-absorption technique in lateral spin valves. *Phys. Rev. B* **98**, 174414 (2018).
39. R. J. Soulen *et al.*, Measuring the Spin Polarization of a Metal with a Superconducting Point Contact. *Science* **282**, 85-88 (1998).
40. M. Althammer *et al.*, Quantitative study of the spin Hall magnetoresistance in ferromagnetic insulator/normal metal hybrids. *Phys. Rev. B* **87**, 224401 (2013).
41. M.-H. Nguyen, D. C. Ralph, R. A. Buhrman, Spin Torque Study of the Spin Hall Conductivity and Spin Diffusion Length in Platinum Thin Films with Varying Resistivity. *Phys. Rev. Lett.* **116**, 126601 (2016).

42. Y. Wang, P. Deorani, X. Qiu, J. H. Kwon, H. Yang, Determination of intrinsic spin Hall angle in Pt. *Appl. Phys. Lett.* **105**, 152412 (2014).
43. S.-J. Wang *et al.*, Long spin diffusion lengths in doped conjugated polymers due to enhanced exchange coupling. *Nat. Electron.* **2**, 98-107 (2019).
44. P. Wang *et al.*, Inverse orbital Hall effect and orbitronic terahertz emission observed in the materials with weak spin-orbit coupling. *npj Quantum Mater.* **8**, 28 (2023).
45. H. Zhang *et al.*, Laser pulse induced efficient terahertz emission from Co/Al heterostructures. *Phys. Rev. B* **102**, 024435 (2020).
46. L. Zhu, D. C. Ralph, R. A. Buhrman, Highly Efficient Spin-Current Generation by the Spin Hall Effect in Au(1-x)P(x). *Phys. Rev. Appl.* **10**, 031001 (2018).
47. M. Kawaguchi, D. Towa, Y.-C. Lau, S. Takahashi, M. Hayashi, Anomalous spin Hall magnetoresistance in Pt/Co bilayers. *Appl. Phys. Lett.* **112**, 152412 (2018).
48. J. Qin, D. Hou, Y. Chen, E. Saitoh, X. Jin, Spin Hall magnetoresistance in Pt/Cr2O3/YIG structure. *J. Magn. Magn. Mater.* **534**, 167980 (2021).
49. D. Go, D. Jo, C. Kim, H.-W. Lee, Intrinsic Spin and Orbital Hall Effects from Orbital Texture. *Phys. Rev. Lett.* **121**, 086602 (2018).
50. D. Go *et al.*, Theory of current-induced angular momentum transfer dynamics in spin-orbit coupled systems. *Phys. Rev. Res.* **2**, 033401 (2020).
51. J. Qiao, J. Zhou, Z. Yuan, W. Zhao, Calculation of intrinsic spin Hall conductivity by Wannier interpolation. *Physical Review B* **98**, 214402 (2018).
52. X. Zhang, S. Li, A. Wang, H. Bao, Pressure-dependent thermal conductivity in Al, W, and Pt: Role of electrons and phonons. *Physical Review B* **106**, 094313 (2022).
53. Y. Wang, P. Deorani, X. Qiu, J. H. Kwon, H. Yang, Determination of intrinsic spin Hall angle in Pt. *Applied Physics Letters* **105**, (2014).
54. J. Zhao *et al.*, Realization of 2D metals at the ångström thickness limit. *Nature* **639**, 354-359 (2025).
55. P. Giannozzi *et al.*, Advanced capabilities for materials modelling with Quantum ESPRESSO. *Journal of Physics: Condensed Matter* **29**, 465901 (2017).
56. P. Giannozzi *et al.*, QUANTUM ESPRESSO: a modular and open-source software project for quantum simulations of materials. *Journal of Physics: Condensed Matter* **21**, 395502 (2009).
57. A. Dal Corso, Pseudopotentials periodic table: From H to Pu. *Computational Materials Science* **95**, 337-350 (2014).
58. N. Marzari, A. A. Mostofi, J. R. Yates, I. Souza, D. Vanderbilt, Maximally localized Wannier functions: Theory and applications. *Reviews of Modern Physics* **84**, 1419-1475 (2012).
59. S. S. Tsirkin, High performance Wannier interpolation of Berry curvature and related quantities with WannierBerri code. *npj Computational Materials* **7**, 1-9 (2021).
60. S. Poncé, E. R. Margine, C. Verdi, F. Giustino, EPW: Electron–phonon coupling, transport and superconducting properties using maximally localized Wannier functions. *Computer Physics Communications* **209**, 116-133 (2016).
61. D. R. Hamann, Optimized norm-conserving Vanderbilt pseudopotentials. *Physical Review B* **88**, 085117 (2013).
62. S. Baroni, S. de Gironcoli, A. Dal Corso, P. Giannozzi, Phonons and related crystal properties from density-functional perturbation theory. *Reviews of Modern Physics* **73**,

515-562 (2001).
63. A. A. Mostofi *et al.*, An updated version of wannier90: A tool for obtaining maximally-localised Wannier functions. *Computer Physics Communications* **185**, 2309-2310 (2014).

# Supporting Information

**P-orbital spin generator with large spin Hall angle and long spin diffusion length**

*Gen Li[1†], Ying Zhang[2,3†], Xiaoguang Xu[1*], Lei Shen[4,3*], Zheng Feng[5], Kangkang Meng[1], Ang Li[6], Lu Cheng[1], Kang He[5], Wei Tan[5], Yong Wu[1], Yihong Wu[2,3], Yong Jiang[1,7*]*

[1] *Key Laboratory of Advanced Materials and Devices for Post-Moore Chips, Ministry of Education, School of Materials Science and Engineering, University of Science and Technology Beijing; Beijing 100083, China.*

[2] *Department of Electrical and Computer Engineering, National University of Singapore; 117576 Singapore.*

[3] *National University of Singapore (Chong Qing) Research Institute; Chongqing Liang Jiang New Area, Chongqing 401123, China.*

[4] *Department of Mechanical Engineering, National University of Singapore; 117575 Singapore.*

[5] *Microsystem & Terahertz Research Center, CAEP; Chengdu 610200, China.*

[6] *Faculty of Materials and Manufacturing, Beijing Key Lab of Microstructure and Properties of Advanced Materials, Beijing University of Technology; Beijing 100124, China.*

[7] *Institute of Quantum Materials and Devices, School of Electronic and Information Engineering; State Key Laboratory of Separation Membranes and Membrane Processes, Tiangong University; Tianjin 300387, China.*

[†]*These authors contributed equally to this work.*

[*]*Corresponding author. Email: xgxu@ustb.edu.cn (X.G.X.); shenlei@nus.edu.sg (L.S.); yjiang@ustb.edu.cn (Y.J.)*

**The PDF file includes:**

Notes S1-S8

Figs. S1-S17

Tables S1 and S2

References and notes

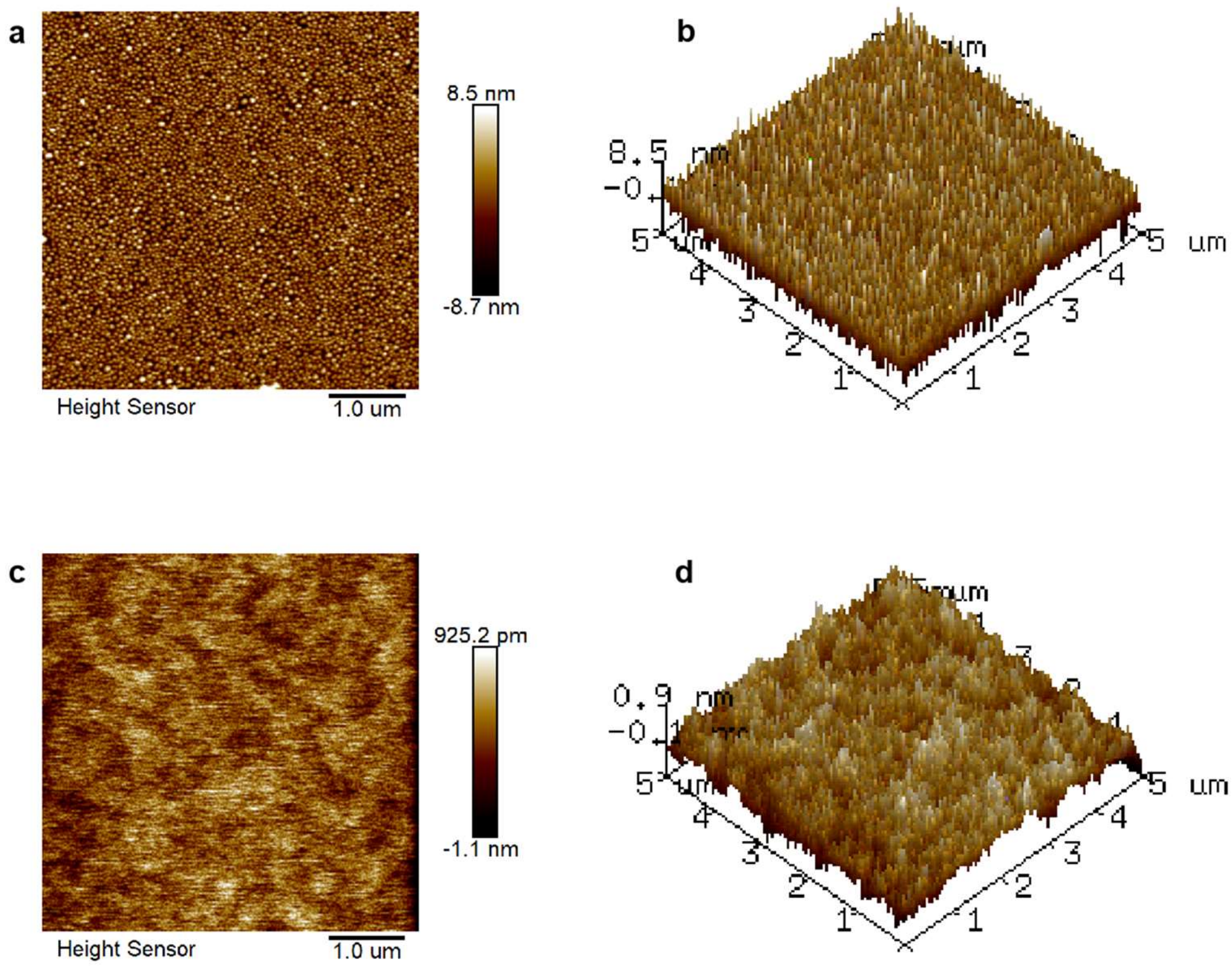

**Figure S1. a**, 2D and **b**, 3D images of a 13 nm $In_2Bi$ film with a roughness $R_q$ of ~2.5 nm. **c**, 2D and **d**, 3D images of a 3 nm Co film with a roughness $R_q$ of ~0.28 nm.

**Note S1: Verification of Electrical Conductivity in $In_2Bi$ Films**

Although the as-deposited $In_2Bi$ films exhibit an insulating behavior due to their granular microstructure, their electrical conductivity can be quantitatively verified by employing a parallel resistance model. we evaluated the resistivity of $In_2Bi$ in Co (3 nm)/$In_2Bi$ (t)/$Al_2O_3$ (3 nm) heterostructures using a parallel resistance model, as the granular $In_2Bi$ film is electrically coupled to the underlying Co layer. As shown in Figure S15, by subtracting the resistance of the Co (3 nm)/$Al_2O_3$ (3 nm) reference device (1527 Ω) and applying the parallel resistance relation, we extracted the effective resistance of $In_2Bi$ as a function of its thickness. The corresponding resistivity was then calculated and found to remain within a relatively narrow range of $5.7 \times 10^{-5}$–$6.7 \times 10^{-5}$ Ω·cm across the studied thickness region, despite the granular morphology. Therefore, using an effective constant value of $6.17 \times 10^{-5}$ Ω·cm for conductivity is a reasonable approximation in Eq. (1), as marked by the red dot in Figure S15. To further confirm the robustness of this assumption, we evaluated the spin Hall parameters at the two extreme resistivity values. For $\rho = 5.7 \times 10^{-5}$ Ω·cm, the extracted spin Hall angle and spin diffusion length are 0.103 and 3.837 nm, respectively. For $\rho = 6.7 \times 10^{-5}$ Ω·cm, the corresponding values are 0.109 and 3.575 nm. These results demonstrate that the variations in resistivity have only a minor effect on the extracted spin Hall parameters, thus validating the treatment of $\rho$ as an effective constant in our SMR analysis.

To further examine whether the SMR signal indeed originates from $In_2Bi$ rather than artifacts related to its granularity, we fabricated continuous $In_2Bi$ films from granular ones using a hot-pressing approach. Although precise thickness control is challenging with this technique, we successfully obtained Co/$In_2Bi$ heterostructures with continuous $In_2Bi$ layers. AFM characterization, as shown in Figures S16a and S16b, confirms a substantial improvement in surface morphology after hot pressing. The roughness of Co (3 nm)/$In_2Bi$ (13 nm) and Co (3 nm)/$In_2Bi$ (30 nm) films are reduced to $R_q = 0.83$ nm and 1.00 nm, respectively, which is significantly lower than that of non-pressed films (Figure S1a). Correspondingly, as displayed in Figures S16c and S16d, these hot-pressed devices exhibit robust SMR signals, substantially larger than that of the Co (3 nm)/$Al_2O_3$ (3 nm) control device ($7.7 \times 10^{-5}$). More importantly, the SMR signal is significantly enhanced after hot pressing. These findings collectively confirm that the observed SMR originates from spin-to-charge interconversion in $In_2Bi$ itself, rather than spurious contributions from granularity.

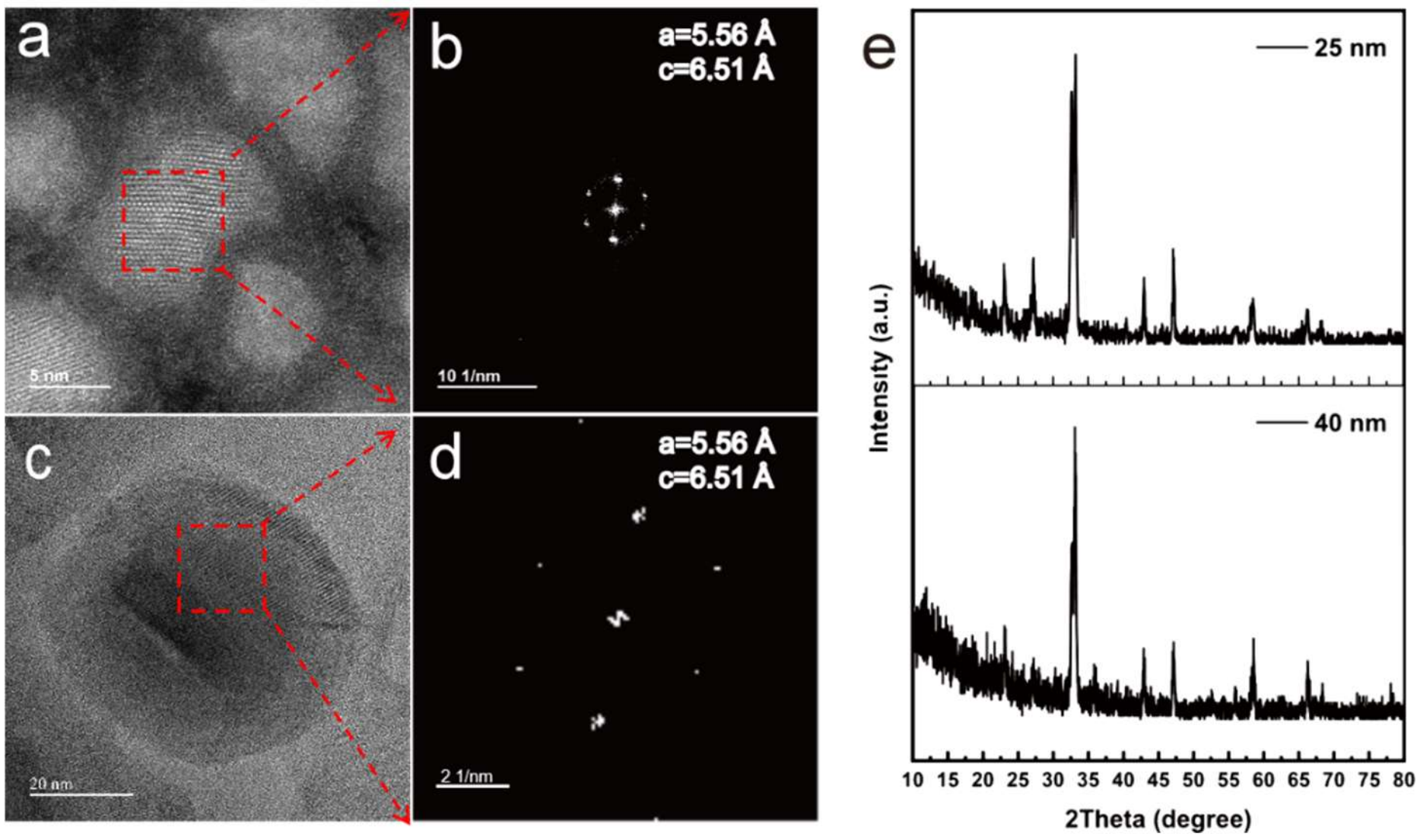

**Figure S2. a,** TEM image and **b,** selected area diffraction pattern of 25 nm $In_2Bi$ single particle. **c,** TEM image and **d,** selected area diffraction pattern of 40 nm $In_2Bi$ single particle. **e,** XRD pattern of $In_2Bi$ film with thickness of 25 nm and 40 nm.

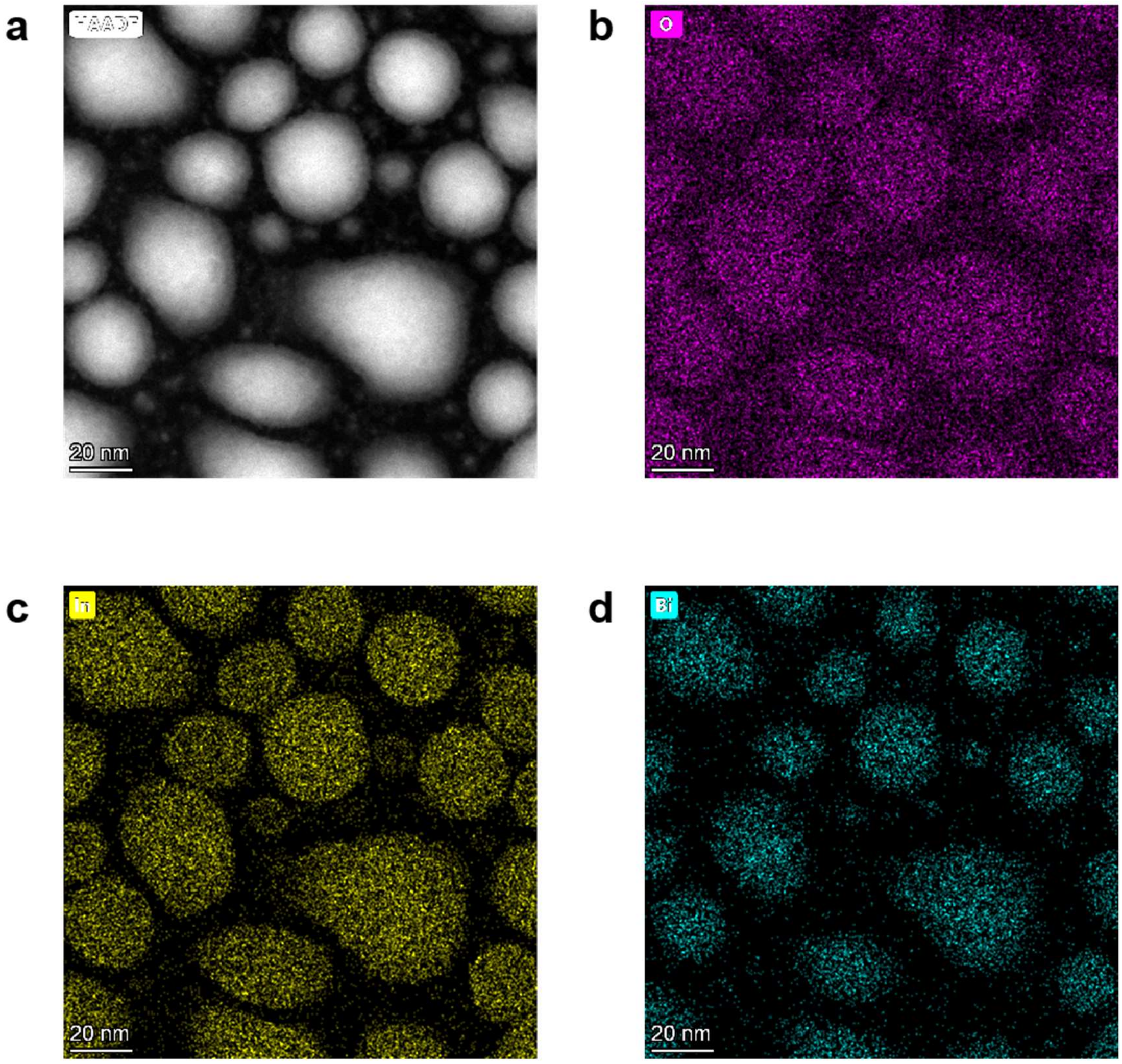

**Figure S3.** **a**, HAADF image of a 13 nm $In_2Bi$. EDS mapping images of **b**, In, **c**, Bi and **d**, O in $In_2Bi$ grown on SiN substrates.

**Note S2: Phase analysis and surface morphology of $In_2Bi$**

The peak areas of XPS spectrum corresponding to In and Bi can be integrated and, using the sensitivity factors (S) for the respective elements, the concentration ratio can be calculated as follows:

$$n_i / n_j = (I_i / S_i) / (I_j / S_j)$$

where $n_{i/j}$, $I_{i/j}$ and $S_{i/j}$ represent the element concentrations, corresponding peak areas, and sensitivity factors, respectively

By accounting for sensitivity factors and integrating the peak intensities, the surface atomic ratio of In/Bi is estimated to be approximately 2.54 using In 3d and Bi 4f orbitals. This value slightly exceeds the target stoichiometry. However, XPS provides only semi-quantitative estimates of atomic ratios. More precise and comprehensive measurement techniques are necessary for an accurate determination of the In/Bi ratio. The intensity ratio of In $d_{5/2}$ to $d_{3/2}$ is approximately 1.527, while the ratio of Bi $f_{7/2}$ to $f_{5/2}$ is about 1.319, aligning with the expected 3:2 and 4:3 ratios for d and f orbitals, respectively.

In order to accurately measure the atomic ratio, inductively coupled plasma optical emission spectrometer (ICP-OES) measurement was employed and demonstrates a precise In/Bi atomic concentration ratio of 2:1, confirming that the resulting $In_2Bi$ films exhibit the desired stoichiometry.

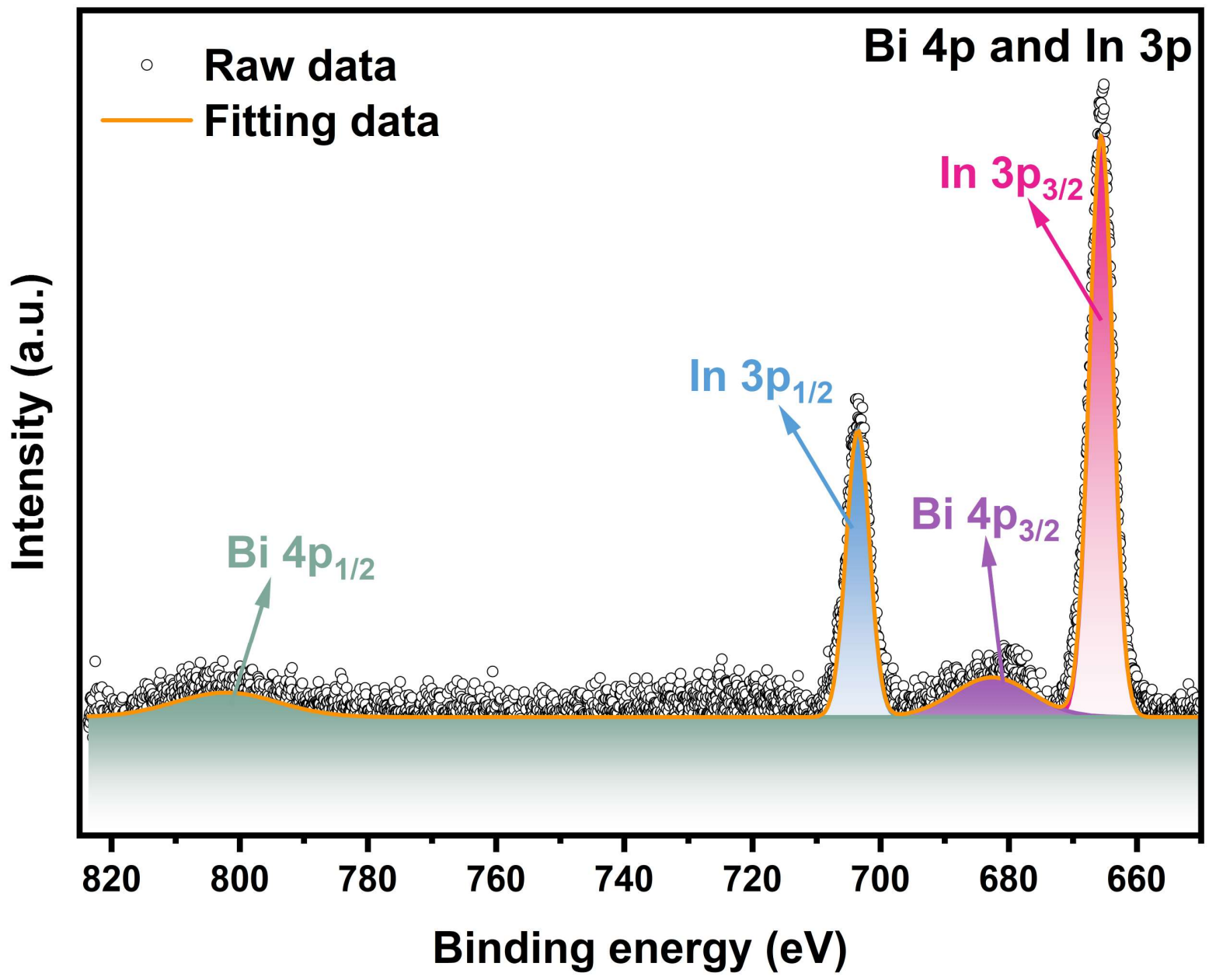

**Figure S4.** XPS spectra of Bi 4p and In 3p orbitals.

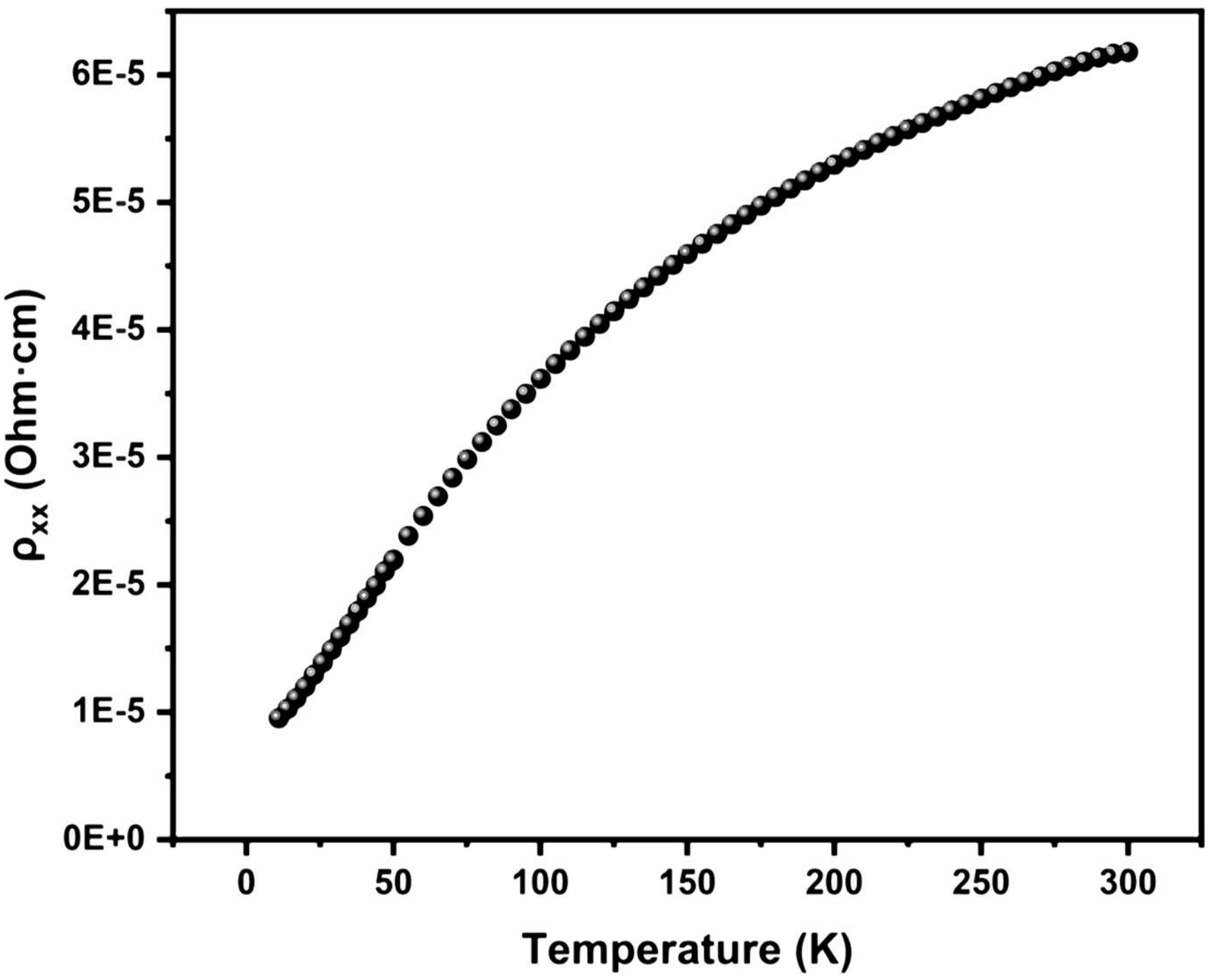

**Figure S5.** $\rho_{xx}$-T curve of $In_2Bi$ with the thickness of 80 nm from 10 K to 300 K.

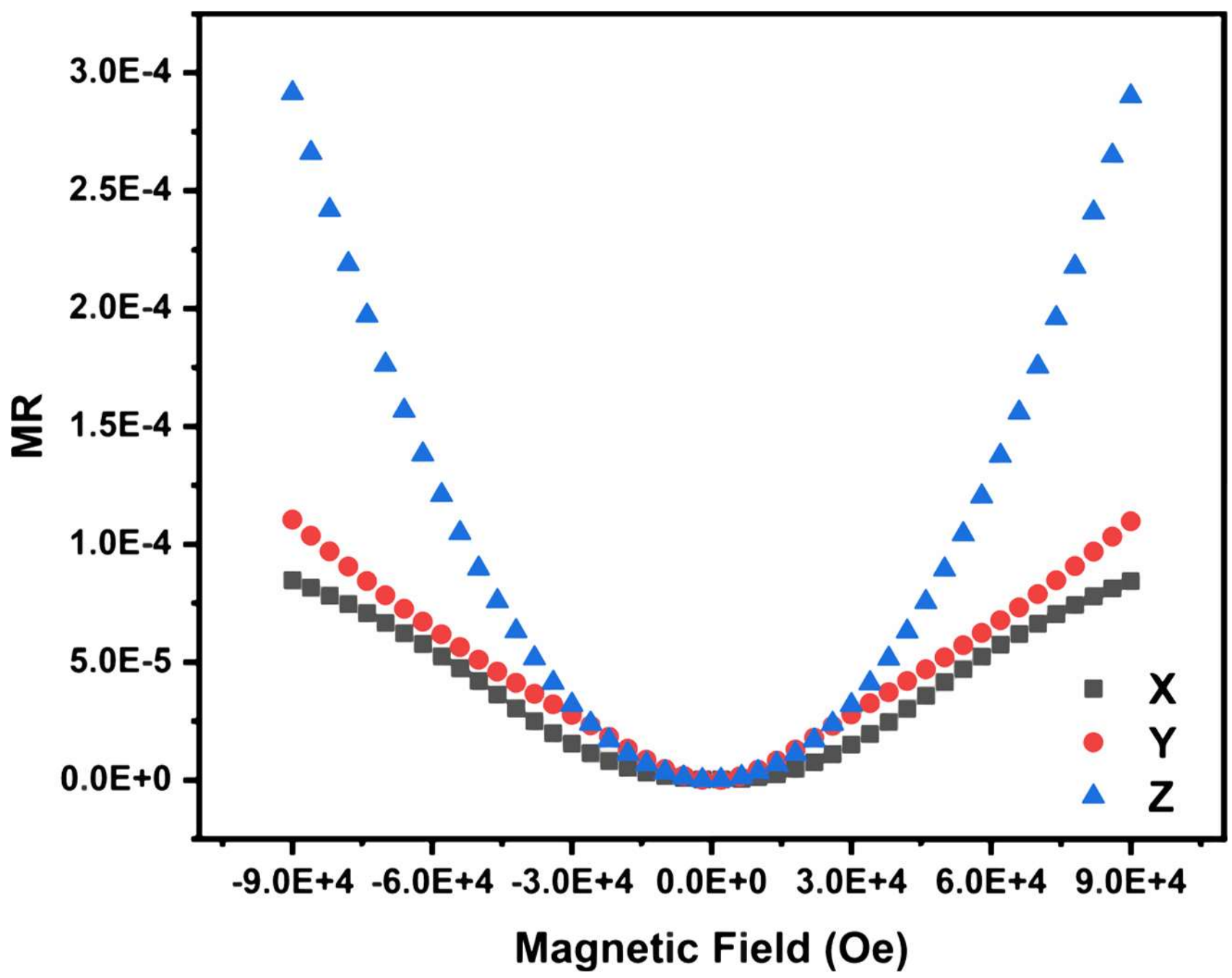

**Figure S6.** Magnetoresistance of 80 nm-$In_2Bi$ film with scanning the magnetic field in X, Y and Z directions.

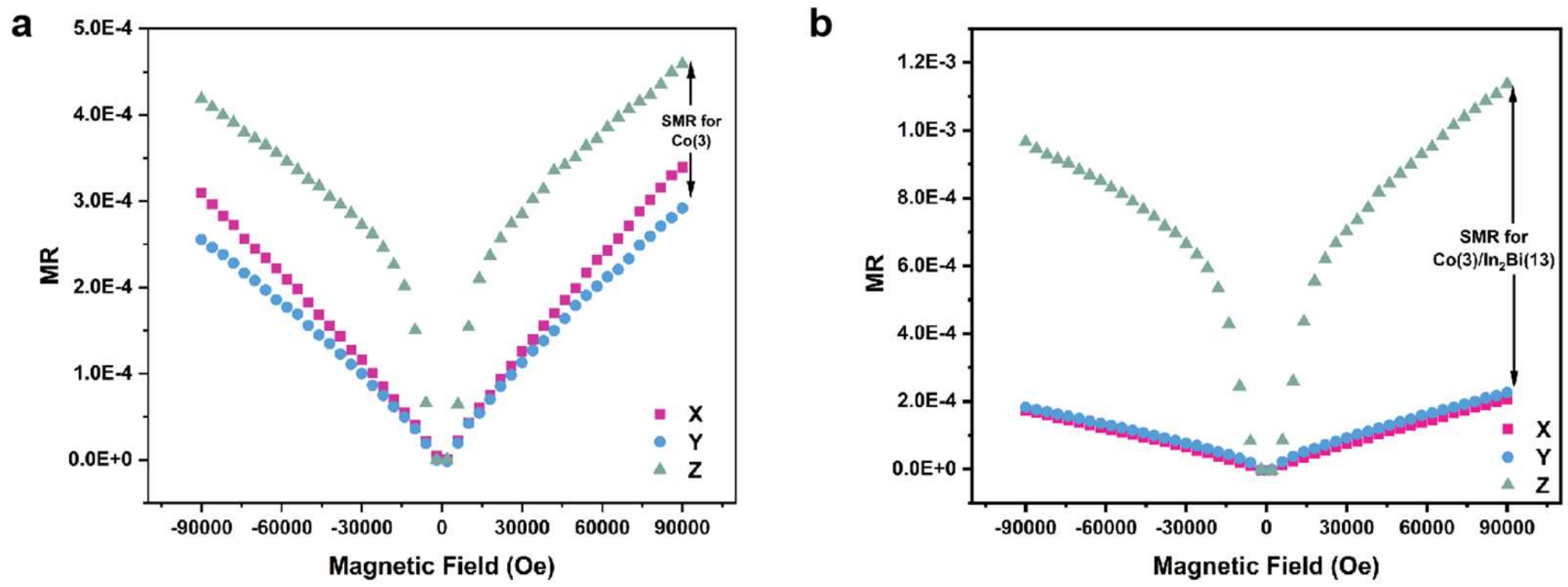

**Figure S7.** Magnetoresistance with scanning the magnetic field in X, Y and Z directions of **a**, Co (3) and **b**, Co (3 nm)/$In_2Bi$ (13 nm).

If there is interfacial alloying between Co and $In_2Bi$, it would likely give rise to additional anisotropic magnetoresistance-like signals. The fact that all three field-sweeping curves exhibit consistent shapes indicates that such extrinsic contributions are negligible. Therefore, we can exclude the influence of possible Co/$In_2Bi$ interfacial alloying on the SMR signal.

**Note S3: The mechanism of spin Hall magnetoresistance**

In SMR testing, both the spin Hall effect (SHE) and the inverse spin Hall effect (ISHE) operate simultaneously. The SHE converts electrical currents into transverse spin currents, while the angle between the spin polarization direction in the non-magnetic layer and the magnetization direction of the adjacent ferromagnetic layer determines whether the spin current is absorbed or reflected. Reflected spin currents are subsequently converted back into charge currents through ISHE, leading to changes in the magnetoresistance (MR).

In this configuration, Co acts as the ferromagnetic layer (FM), $In_2Bi$ as the non-magnetic layer (NM), and $Al_2O_3$ as the capping layer. The SHE in $In_2Bi$ induces a spin current $J_S$ , which propagates perpendicularly to the film surface with a spin polarization direction parallel to the interface. As the device rotates in the yz-plane, the magnetization M remains aligned with the external magnetic field. When the spin polarization and M are not collinear, a portion of the spin current is absorbed by the magnetization at the NM/FM interface, while another portion is reflected and reconverted into charge current via ISHE. When M is parallel to the spin polarization, absorption is zero, resulting in maximum $J_e$. When M is perpendicular to the spin polarization, spin current is absorbed by FM layer, corresponding to minimu m $J_e$. Thus, the direction of magnetization in Co affects the resistance of the $In_2Bi$ film, yielding SMR.

**Note S4: SMR fitting of $In_2Bi$ film with different thickness**

We performed SMR measurements on Co (3 nm)/ $In_2Bi$ (t) heterostructures, with $In_2Bi$ layer thicknesses of t = 0, 3, 5, 7, 9, 11, 13, 15, 17, 25, and 40 nm. When the spin polarization in the $In_2Bi$ layer is parallel to the magnetization of the Co layer, spin current is reflected at the interface and reconverted into charge current via the ISHE in the $In_2Bi$ layer. This additional charge current adds to the original longitudinal charge current, resulting in a low-resistance state. Conversely, when the spin polarization is perpendicular to the magnetization, the spin current is absorbed by the Co layer at the interface, suppressing reflection and leading to a reduced contribution to the longitudinal current, manifesting as a high-resistance state.

The SMR can be described by the following relationship:

$$\rho_{xx} = \rho_0 + \Delta\rho_{xx}[\hat{m} \cdot (\hat{j} \times \hat{z})]^2 \tag{S1}$$

where $\rho_0$ is the normal resistivity, $\rho_{xx}$ is the amplitude of the SMR, and $\hat{m}$、$\hat{j}$、$\hat{z}$ represent the magnetization, current direction, and the vector normal to the interface, respectively.

In the y-z plane, as the magnetization rotates under the external magnetic field, it remains perpendicular to the current direction. Thus, anisotropic magnetoresistance (AMR) has no contribution to MR, and only SMR is taken into account. The angular dependence of SMR, observed as the magnetic field rotates in the plane, is depicted in Fig S8.

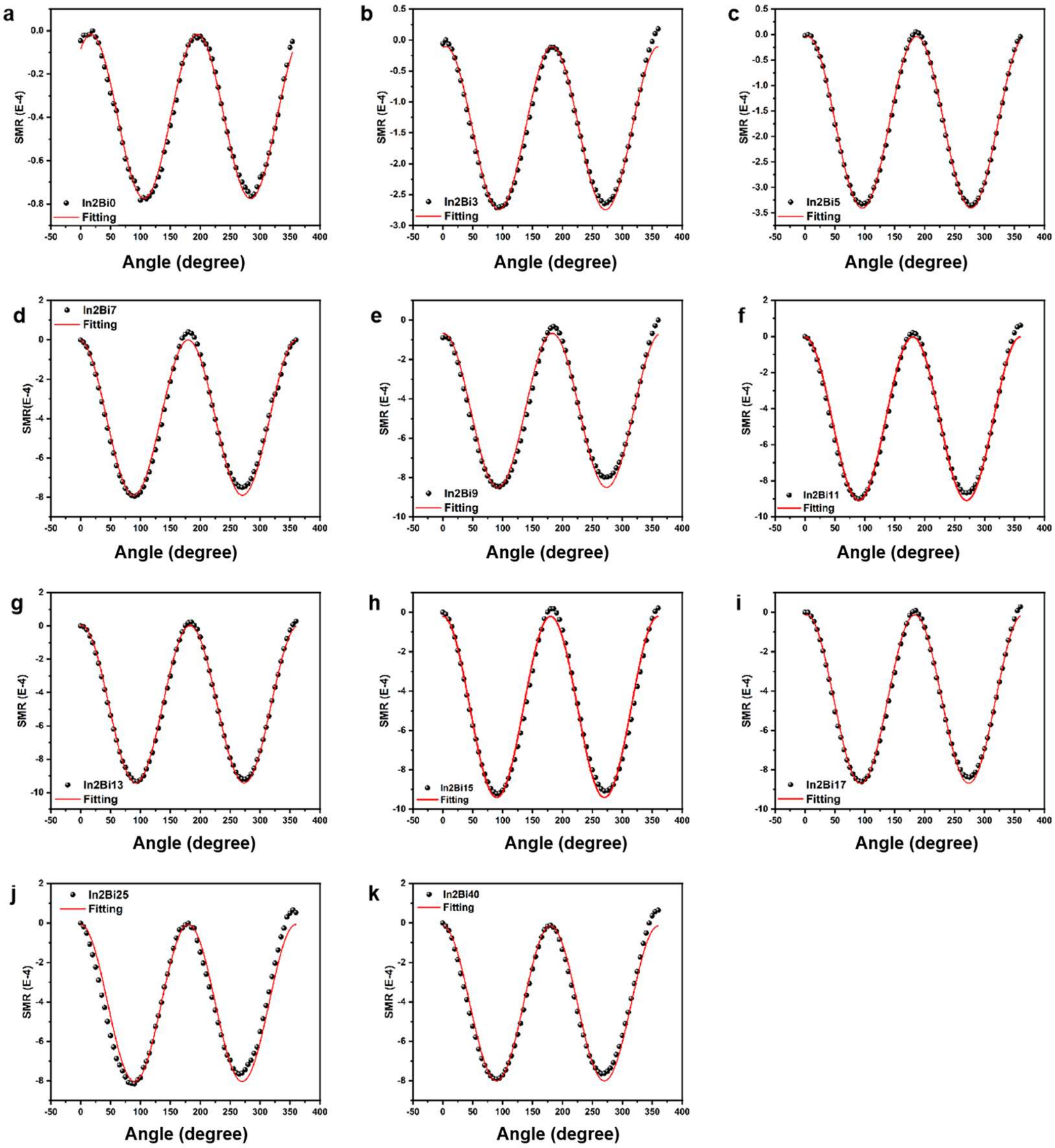

**Figure S8. a-k**, SMR curves of Co (3 nm)/$In_2Bi$ (t) in different $In_2Bi$ thickness, t= 0 nm, 3 nm, 5 nm, 7 nm, 9 nm, 11 nm, 13 nm, 15 nm, 17 nm, 25 nm and 40 nm.

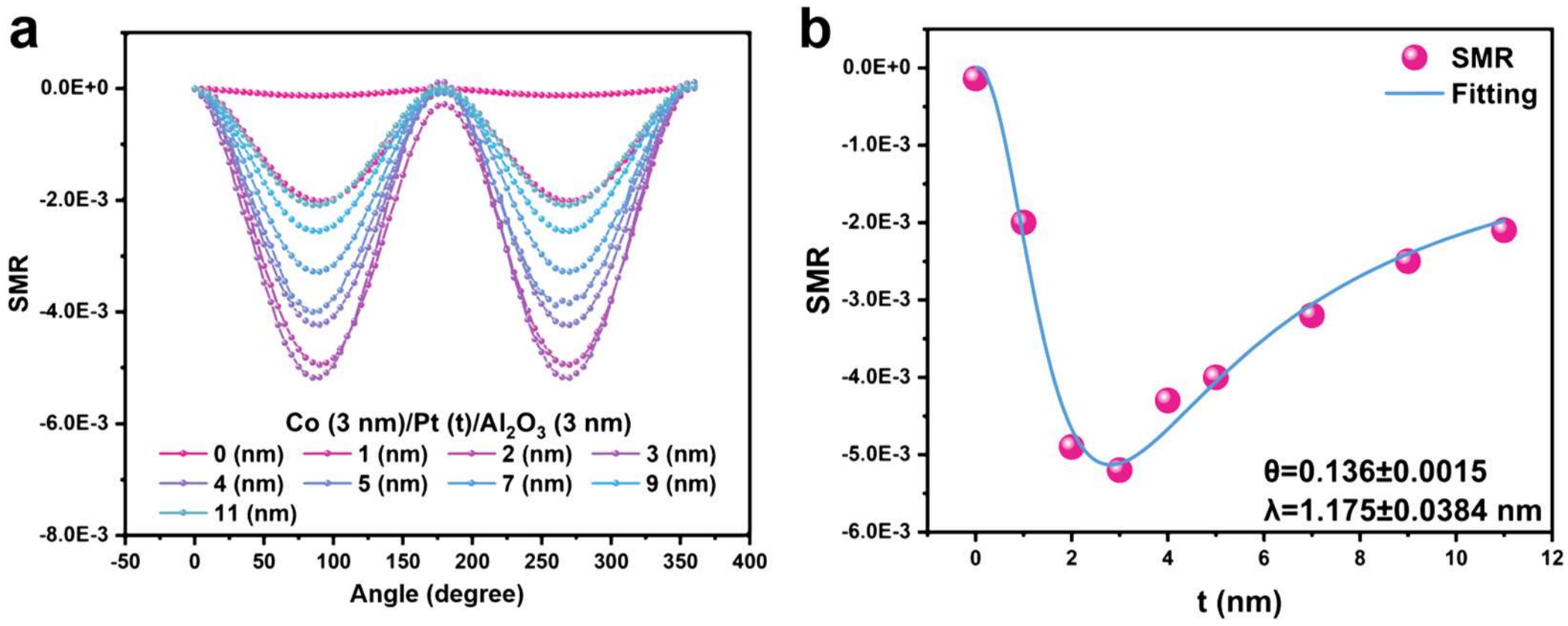

**Figure S9. a,** Angle-dependent spin magnetoresistance curves of Co (3 nm)/Pt (t)/$Al_2O_3$ (3 nm) devices with varying Pt thickness measured under 9 T at room temperature. **b,** Dependence of the SMR amplitude on the Pt thickness in Co (3 nm)/Pt (t)/$Al_2O_3$ (3 nm) heterojunction and the fitting curve.

## Note S5: THz emission test mechanism

The granular nature of the $In_2Bi$ film may influence conventional electrical transport measurements. Consequently, an alternative characterization method is necessary to validate the spin currents generated in $In_2Bi$, separate from traditional electrical transport testing. The ISHE has recently emerged as a pivotal factor in extending the field of spintronics into the terahertz (THz) domain *(S1)*. THz spintronics holds considerable promise for applications in high-speed currents and computing technologies *(S2)*. Therefore, THz emission test was used to further verify the spin current in $In_2Bi$. In the context of THz emission induced by ISHE in ferromagnetic/non-magnetic (FM/NM) layers, a femtosecond laser pulse pumps the FM/NM heterostructure, generating non-equilibrium spin-polarized electrons within the FM layer. These electrons subsequently diffuse through the non-magnetic layer via super diffusion *(S3, S4)*. Due to ISHE in the NM layer, the spin current is converted into a transient transverse charge current, which produces short THz pulses that propagate perpendicularly to the current direction.

**Table S1: Different thicknesses in NM layers and corresponding THz signal intensities**

| Co (3 nm)/$In_2Bi$ (t) | | Co (3 nm)/Pt (t) | |
|---|---|---|---|
| **Thickness (nm)** | **THz Signal (a.u.)** | **Thickness (nm)** | **THz Signal (a.u.)** |
| **0** | 3.8 | **0** | 10.2 |
| **3** | 84.7 | **1** | 190.9 |
| **5** | 163.8 | **2** | 856.8 |
| **7** | 218.6 | **3** | 898.0 |
| **9** | 260.8 | **4** | 834.5 |
| **11** | 280.0 | **5** | 736.0 |
| **13** | 299.0 | **7** | 557.3 |
| **15** | 287.2 | **10** | 345.1 |
| **17** | 282.0 | **13** | 202.4 |
| **19** | 264.8 | **16** | 138.3 |
| **30** | 237.6 | | |
| **40** | 178.6 | | |

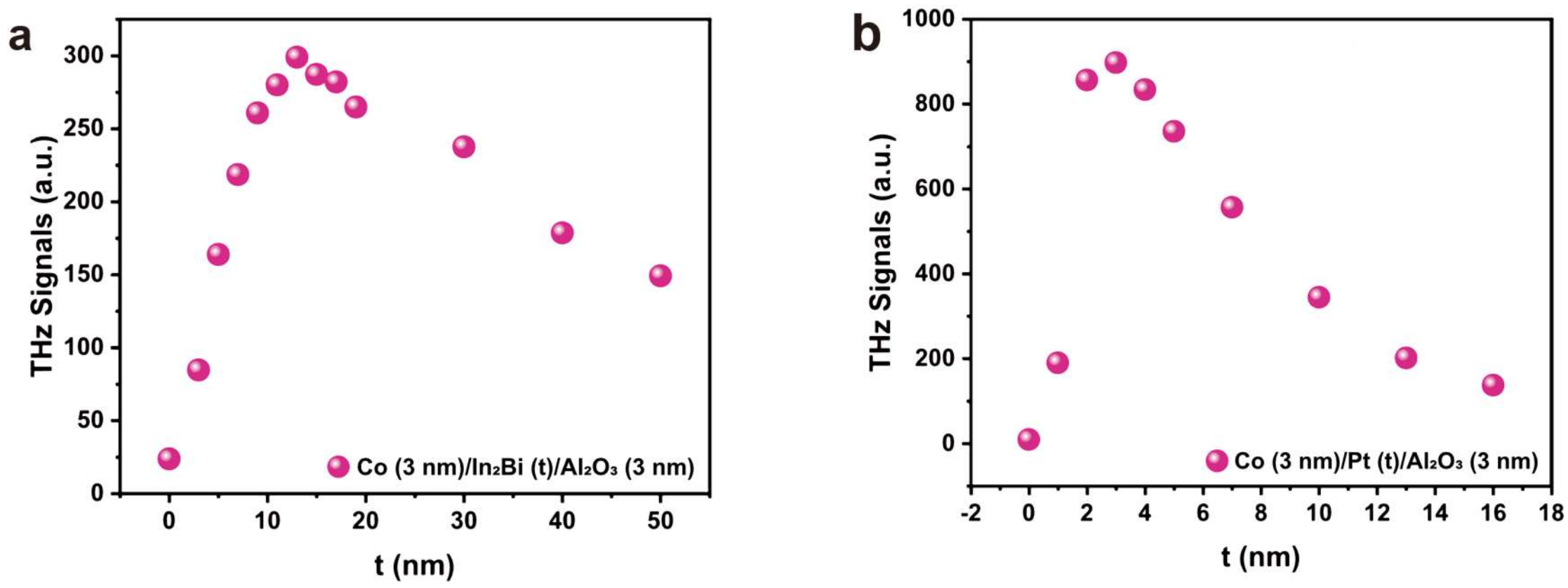

**Figure S10. a.** The $In_2Bi$ thickness dependence of the THz signals for Co (3 nm)/$In_2Bi$ (t)/$Al_2O_3$ (3 nm). **b.** The Pt thickness dependence of the THz signals for Co (3 nm)/Pt (t)/$Al_2O_3$ (3 nm).

**Note S6: THz emission test of Co/Pt heterostructure**

Co/Pt heterostructures were obtained under the same experimental conditions as those employed for the Co/$In_2Bi$ stack. Co (3 nm)/Pt (t) heterostructures (where t = 0, 1, 2, 3, 4, 5, 7, 10, 13 and 16 nm) are fabricated and conducted THz measurements. Just like Co/$In_2Bi$ samples, the signal strength is quantified as the difference between the amplitudes of peaks and troughs. The results were obtained under the same experimental conditions as those employed for the Co/$In_2Bi$ stack. The signal strength reached a peak of approximately 900. The results demonstrate that as the thickness of Pt increases, the signal strength in the Co/Pt heterojunction initially rises and then declines. Similar to the observations in SMR, the THz intensity also reach peaks at the Pt thickness reaches 2-3 nm. The fitting results indicate that $\lambda_{Pt}$=2.93±0.24 nm.

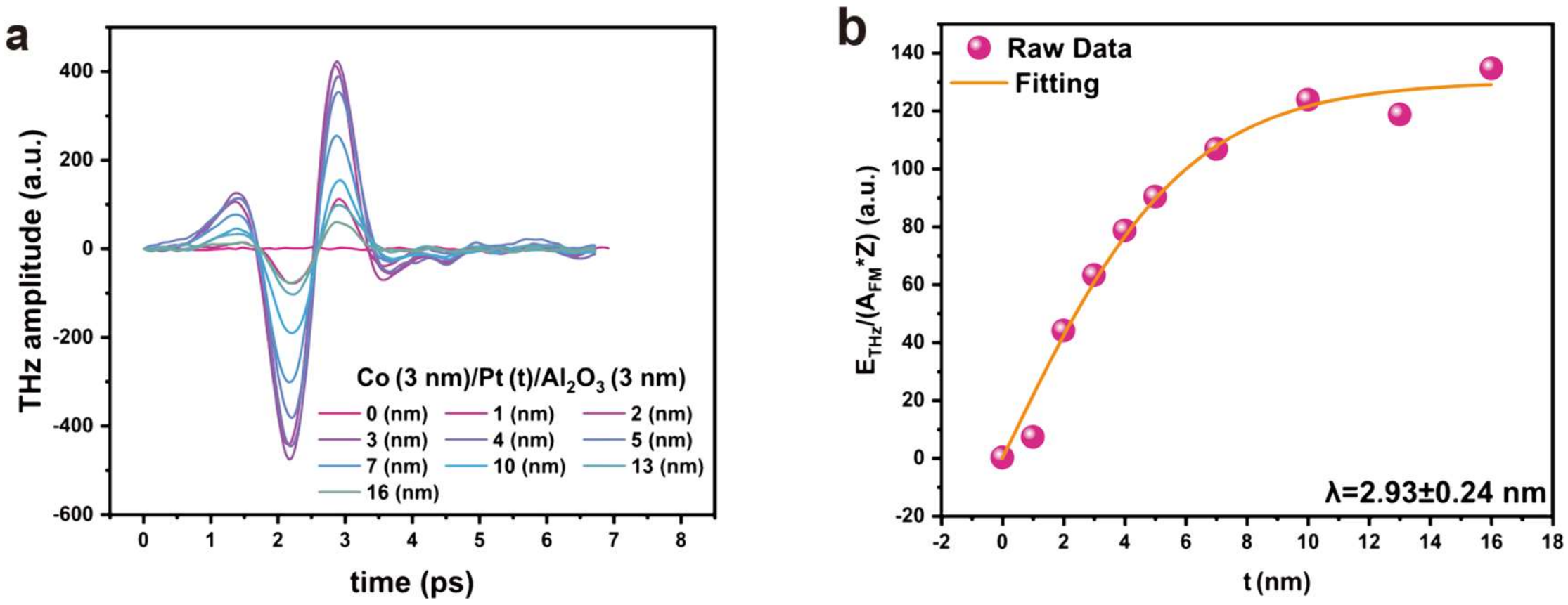

**Figure S11. a**, THz emission test of Co (3 nm)/Pt (t)/$Al_2O_3$ (3 nm) sample. **b**, The Pt thickness dependence of the THz amplitude for Co (3 nm)/Pt (t)/$Al_2O_3$ (3 nm) spintronic THz emitter. The symbols and lines reveal the experimental and theoretical results, respectively.

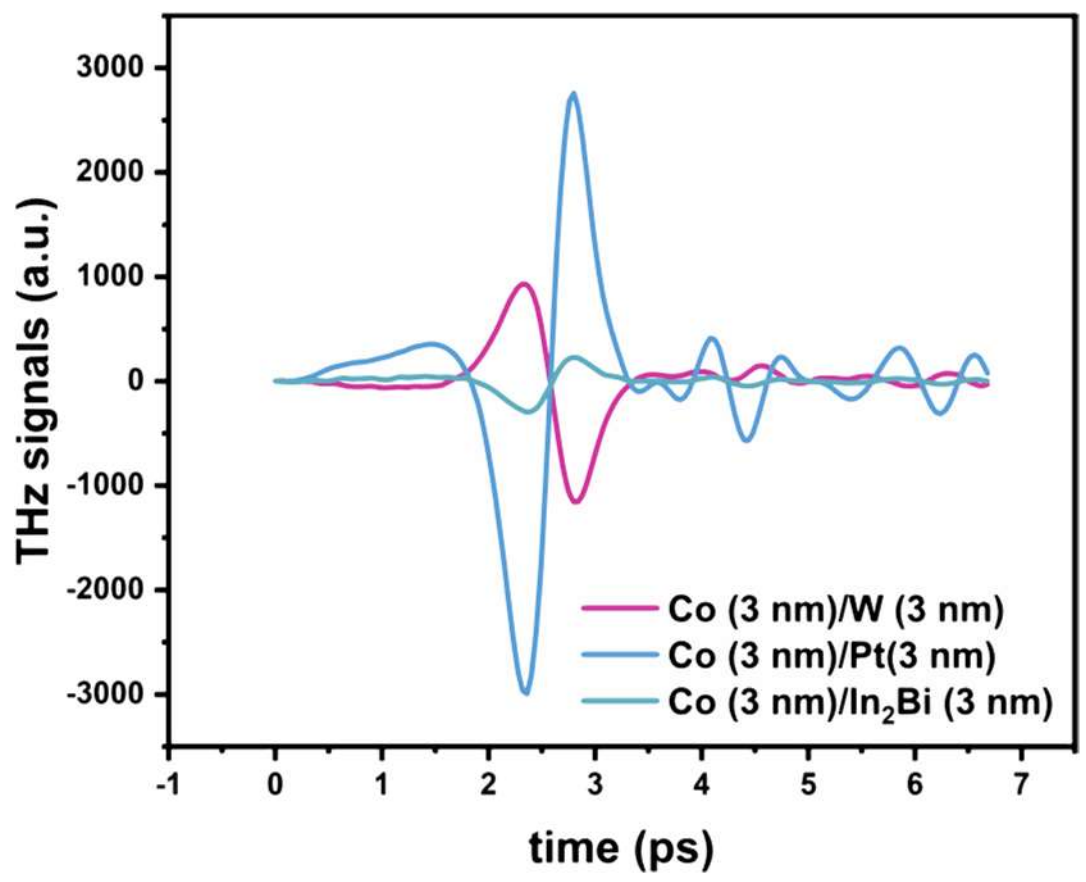

**Figure S12.** THz emission signals of Co (3 nm)/W (3 nm), Co (3 nm)/Pt (3 nm) and Co (3 nm)/$In_2Bi$ (3 nm) heterojunctions measured under the same conditions. The results show that the spin current direction of $In_2Bi$ is the same as that of Pt but opposite to that of W.

**Table S2. The SHA and SDL of Pt obtained by different test methods**

| Method | SHA | SDL (nm) | Reference |
|---|---|---|---|
| SMR | 0.11 ± 0.08 | 1.5±0.5 | *(40)* |
| SMR | 0.136 | 1.175±0.04 | This work |
| THz emission | - | 4±0.2 | *(S5)* |
| THz emission | - | 2.93±0.24 | This work |
| Spin absorption measurements | 0.017-0.022 | 6.5-10.1 | *(S6)* |
| Spin torque efficiency measurements | - | 5.1±0.5 | *(S7)* |
| Spin pumping | 0.027±0.005 | ~4 | *(S8)* |
| ST-FMR | 0.068 | ~1.5 | *(42)* |

**Note S7: Crystal structure and SHC of $In_2Bi$ in phase (102) and (110)**

In order to analyze the SHC of $In_2Bi$ in different phases, we construct $In_2Bi$ structures in phases (102) and (110), and took 4 and 6 layers of atoms as examples to simulate the actual performance of $In_2Bi$ materials under experimental conditions.To facilitate analysis, we listed the complete SHC tensor of bulk $In_2Bi$ and bulk Pt for comparasion.

The SHC tensor of bulk $In_2Bi$ at fermi energy[1]:

$$\sigma^x = \begin{matrix} 0 & 0 & 0 \\ 0 & 0 & -1058.11 \\ 0 & 1109.90 & 0 \end{matrix} \; (\hbar/e(S/cm))$$

$$\sigma^y = \begin{matrix} 0 & 0 & 1058.59 \\ 0 & 0 & 0 \\ -1110.38 & 0 & 0 \end{matrix} \; (\hbar/e(S/cm))$$

$$\sigma^z = \begin{matrix} 0 & -1456.13 & 0 \\ 1455.27 & 0 & 0 \\ 0 & 0 & 0 \end{matrix} \; (\hbar/e(S/cm))$$

The SHC tensor of bulk Pt at fermi energy[1]:

$$\sigma^x = \begin{matrix} 0 & 0 & 0 \\ 0 & 0 & -2234.81 \\ 0 & 2234.81 & 0 \end{matrix} \; (\hbar/e(S/cm))$$

$$\sigma^y = \begin{matrix} 0 & 0 & 2234.81 \\ 0 & 0 & 0 \\ -2234.81 & 0 & 0 \end{matrix} \; (\hbar/e(S/cm))$$

$$\sigma^z = \begin{matrix} 0 & -2234.81 & 0 \\ 2234.81 & 0 & 0 \\ 0 & 0 & 0 \end{matrix} \; (\hbar/e(S/cm))$$

Extensive studies have shown that the crystal phase of the protected nodal lines contribute significantly to the larger SHC, so the symmetry of the structure in different phases are fully analyzed. Protected by six-fold symmetry and multiple crystal plane symmetry, the original $In_2Bi$ lattice has 24 symmetric operations, while the $In_2Bi$ of (102) retains only one mirror symmetry plane. The overall lattice structure presents a high degree of disorder, resulting in a significant decline in the calculated SHC (Figure

S13). Symmetry breaking results in the apperance of more than three non-zero components in the SHC tensor. Although the positions of these non-zero terms in the tensor show a regular arrangement, there is no inverse sign phenomenon consistent with the bulk structure.

**SHC tensor for $In_2Bi$ in (102) phase with 4-atom-layer at fermi energy[1]:**

$$\sigma^x = \begin{matrix} 0 & 57.69 & 0 \\ 64.44 & 0 & 8.80 \\ 0 & 40.63 & 0 \end{matrix} \quad (\hbar/e(S/cm))$$

$$\sigma^y = \begin{matrix} -55.09 & 0 & 31.39 \\ 0 & 147.42 & 0 \\ -32.73 & 0 & 5.37 \end{matrix} \quad (\hbar/e(S/cm))$$

$$\sigma^z = \begin{matrix} 0 & -276.29 & 0 \\ 281.14 & 0 & 7.57 \\ 0 & -1.89 & 0 \end{matrix} \quad (\hbar/e(S/cm))$$

**SHC tensor for $In_2Bi$ in (102) phase with 6-atom-layer at fermi energy[1]:**

$$\sigma^x = \begin{matrix} 0 & -58.83 & 0 \\ 130.07 & 0 & -76.46 \\ 0 & 77.56 & 0 \end{matrix} \quad (\hbar/e(S/cm))$$

$$\sigma^y = \begin{matrix} -62.18 & 0 & 136.68 \\ 0 & 7.75 & 0 \\ -69.94 & 0 & 3.67 \end{matrix} \quad (\hbar/e(S/cm))$$

$$\sigma^z = \begin{matrix} 0 & -353.61 & 0 \\ 516.58 & 0 & -120.88 \\ 0 & 9.36 & 0 \end{matrix} \quad (\hbar/e(S/cm))$$

As for the $In_2Bi$ in (110) phase (Figure S14), the symmetry of the original $In_2Bi$ structure is more preserved, thus retaining relatively higher SHC than (102) phase. Compared with the 4-atom-layer structure, the 6-atom-layer structure is closer to the actual bulk structure, so the calculated results are closer to the value of the bulk system. In addition, the preservation of symmetry is also reflected in the SHC tensor. The tensor on the (110) plane maintains consistency with the original bulk structure, as evidenced by the number, positions, and directional relationships of its non-zero components aligning with those of the bulk. However, due to the reduction of symmetry, the inverse sign relationship between the elements is not fully preserved. The construction of the cleaved plane reduces the spin polarization components along the x and y directions to one-tenth of their bulk values, while the z-direction component is approximately halved.

[1] For each element in SHC tensor, the superscripts denote the spin polarization direction, while the subscripts represent the spin current direction and charge current direction, respectively.

**SHC tensor for $In_2Bi$ in (110) phase with 4-atom-layer at fermi energy[1]:**

$$\sigma^x = \begin{matrix} 0 & 0 & 0 \\ 0 & 0 & -87.54 \\ 0 & 78.64 & 0 \end{matrix} \quad (\hbar/e(S/cm))$$

$$\sigma^y = \begin{matrix} 0 & 0 & 64.46 \\ 0 & 0 & 0 \\ -89.85 & 0 & 0 \end{matrix} \quad (\hbar/e(S/cm))$$

$$\sigma^z = \begin{matrix} 0 & -488.81 & 0 \\ 799.62 & 0 & 0 \\ 0 & 0 & 0 \end{matrix} \quad (\hbar/e(S/cm))$$

**SHC tensor for $In_2Bi$ in (110) phase with 6-atom-layer at fermi energy[1]:**

$$\sigma^x = \begin{matrix} 0 & 0 & 0 \\ 0 & 0 & -179.97 \\ 0 & 133.05 & 0 \end{matrix} \quad (\hbar/e(S/cm))$$

$$\sigma^y = \begin{matrix} 0 & 0 & 174.00 \\ 0 & 0 & 0 \\ -153.66 & 0 & 0 \end{matrix} \quad (\hbar/e(S/cm))$$

$$\sigma^z = \begin{matrix} 0 & -742.97 & 0 \\ 674.18 & 0 & 0 \\ 0 & 0 & 0 \end{matrix} \quad (\hbar/e(S/cm))$$

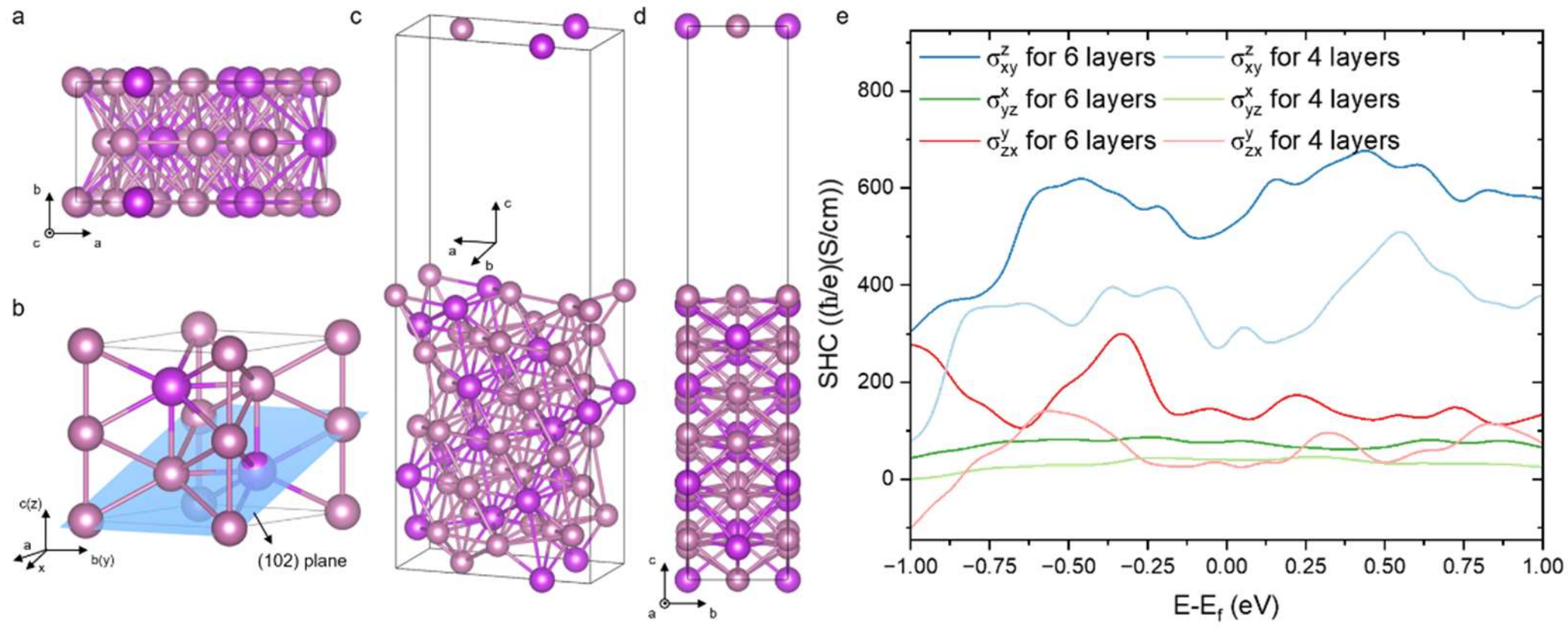

**Figure S13.** Crystal structure and SHC of $In_2Bi$ in phase (102). **a-d**, The crystal structure of $In_2Bi$ in phase (102) with 4-atom-layer. **e**, Three main non-zero quantities in the SHC tensor plotted as a function of energy. The high and low saturation color groups represent the results of 6-layer and 4-layer structure, respectively.

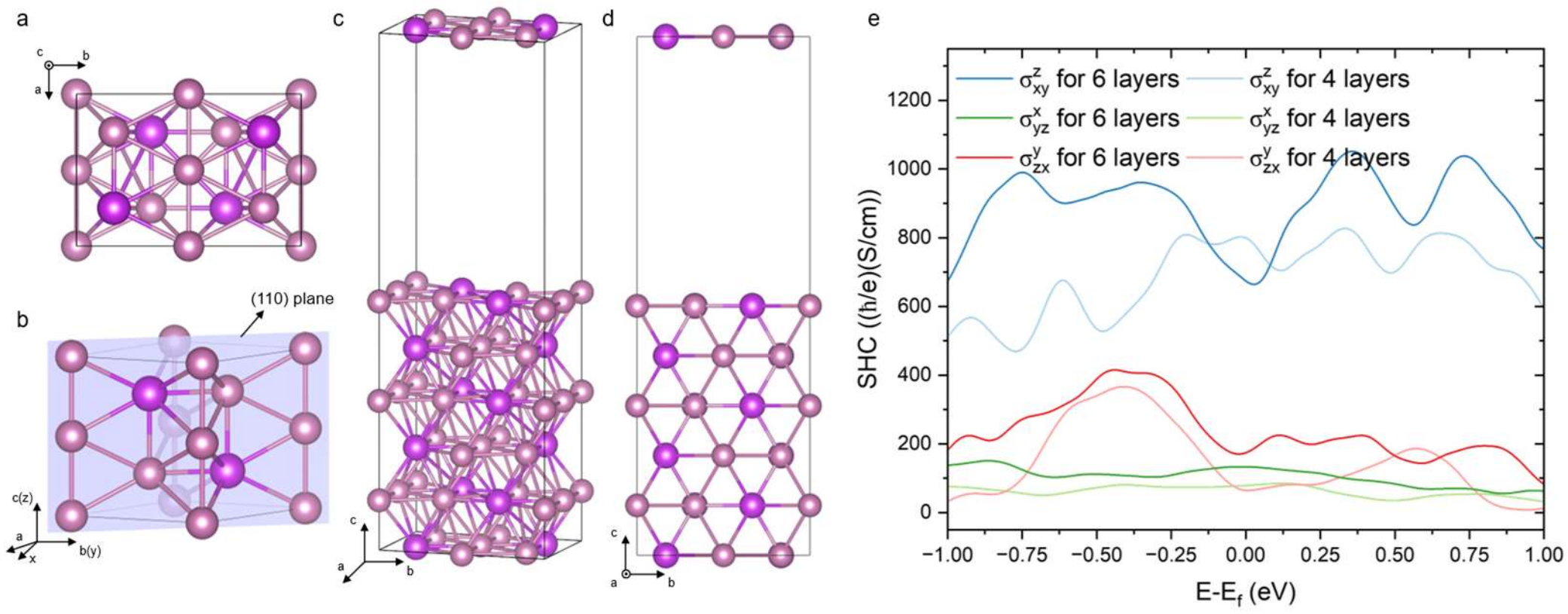

**Figure S14.** Crystal structure and SHC of $In_2Bi$ in phase (110). **a-d**, The crystal structure of $In_2Bi$ in phase (110) with 4-atom-layer. **e**, Three main non-zero quantities in the SHC tensor plotted as a function of energy. The high and low saturation color groups represent the results of 6-layer and 4-layer structure, respectively.

**Note S8: Method comparison and k-mesh converge test**

In order to obtain more accurate SHC calculation results while taking into account the computational amount, we take the $\sigma_{xy}^{z}$ component of SHC for $In_2Bi$ bulk material as an example to test the calculation performance of Wannier90 and WannierBerri in different k-point grids (Figure S17). With the increase of the k-point grid, the calculated values of the two methods gradually tend to the same curve, which reflects the consistency of the results under different algorithms. The calculation accuracy and calculation speed of Wannier90 are greatly affected by the k-point grid. The coarser grid makes the results more volatile and significantly underestimates the value, while the denser grid will significantly increase the calculation time. WannierBerri leverages the symmetry properties of crystals, requiring only irreducible k-points for computation. This approach significantly reduces the computational workload. Aside from noticeable fluctuations in results when the k-point mesh size is set to 30, the calculations demonstrate high consistency with further increases in the k-point density.

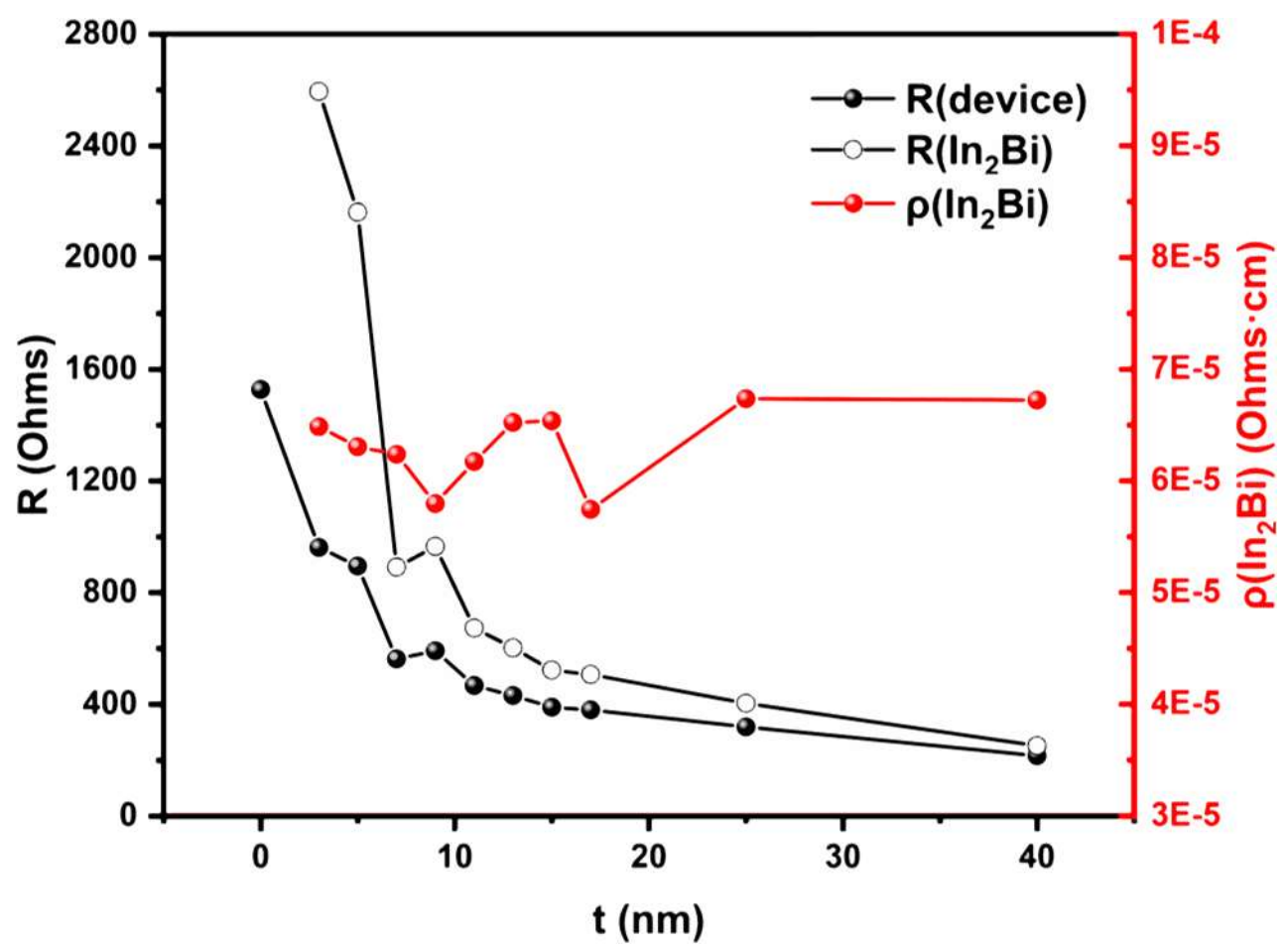

**Figure S15.** Thickness dependence of resistance and resistivity of $In_2Bi$ films. Device resistance $R_{(device)}$ (black solid line), extracted $In_2Bi$ resistance $R_{(In_2Bi)}$ (black hollow line) and corresponding resistivity $\rho_{(In_2Bi)}$ (red solid line) as a function of $In_2Bi$ thickness.

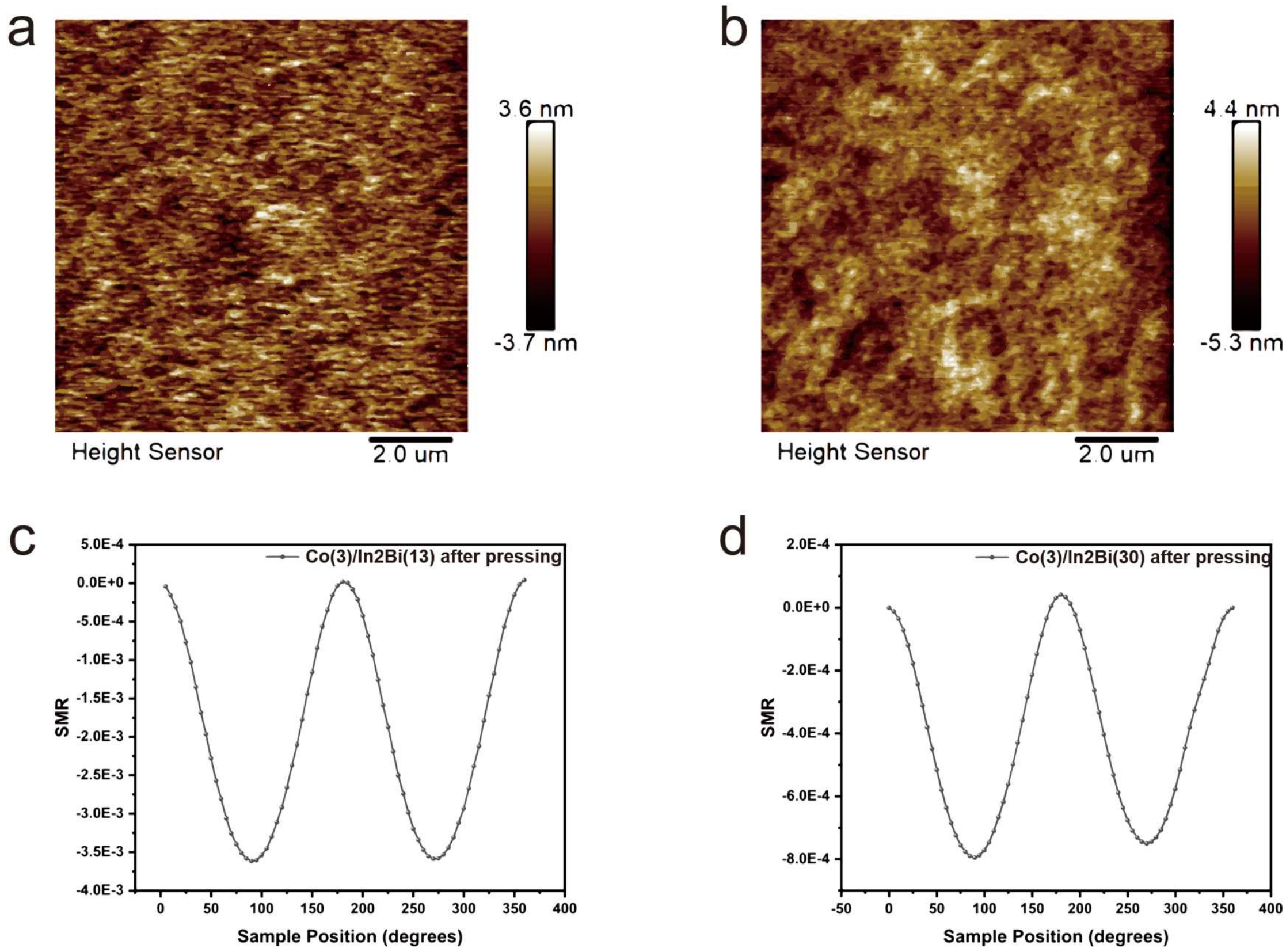

**Figure S16. a,** AFM images of a Co (3 nm)/$In_2Bi$ (13 nm) film after hot-pressing with a roughness Rq of 0.83 nm. **b,** AFM images of a Co (3 nm)/$In_2Bi$ (30 nm) film after hot-pressing with a roughness Rq of 1.00 nm. **c,** SMR curves of Co (3 nm)/$In_2Bi$ (13 nm) film after pressing. **d,** SMR curves of Co (3 nm)/$In_2Bi$ (30 nm) film after pressing.

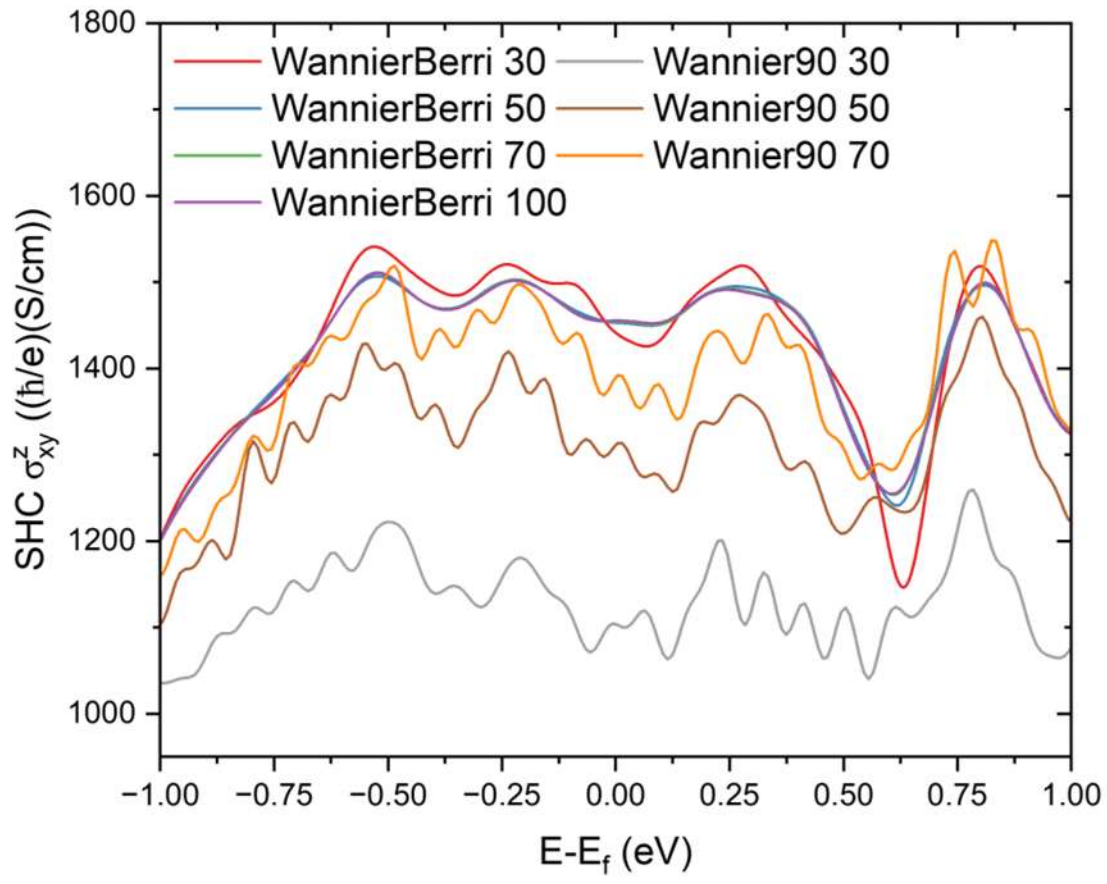

**Figure S17.** Comparison of two calculated method and corresponding k-mesh converge test. The legend indicates the calculation method first, and then the corresponding k-point grid.

## REFERENCES AND NOTES

(S1) T. Seifert et al., Efficient metallic spintronic emitters of ultrabroadband terahertz radiation. Nat. Photonics **10**, 483-488 (2016).

(S2) J. Walowski, M. Münzenberg, Perspective: Ultrafast magnetism and THz spintronics. J. Appl. Phys. **120**, 140901, (2016).

(S3) M. Battiato, K. Carva, P. M. Oppeneer, Superdiffusive Spin Transport as a Mechanism of Ultrafast Demagnetization. Phys. Rev. Lett. **105**, 027203 (2010).

(S4) A. Melnikov et al., Ultrafast Transport of Laser-Excited Spin-Polarized Carriers in Au/Fe/MgO (001). Phys. Rev. Lett. **107**, 076601 (2011).

(S5) Y. Wang, W. Li, H. Cheng, Z. Liu, Z. Cui, J. Huang, B. Xiong, J. Yang, H. Huang, J. Wang, et al. Enhancement of spintronic terahertz emission enabled by increasing Hall angle and interfacial skew scattering. Commun. Phys., **6**, 280 (2023)

(S6) E. Sagasta, Y. Omori, M. Isasa, M. Gradhand, L. E. Hueso, Y. Niimi, Y. Otani, and F. Casanova. Tuning the spin Hall effect of Pt from the moderately dirty to the superclean regime. Phys. Rev. B, 060412 (2016).

(S7) M. H. Nguyen, D. C. Ralph, and R. A. Buhrman. Spin Torque Study of the Spin Hall Conductivity and Spin Diffusion Length in Platinum Films with Varying Resistivity. Phys. Rev. Lett. **116**, 126601 (2016).

(S8) V. Vlaminck, J. E. Pearson, S. D. Bader, and A. Hoffman. Dependence of spin-pumping spin Hall effect measurements on layer thickness and stacking order. Phys. Rev. B **88**, 064414 (2013).